\documentclass[intlimits,twoside,a4paper]{article}

\usepackage[cp1251]{inputenc}
\usepackage[]{amsmath}

\usepackage[eqsecnum]{cmpj3}

\issue{2024}{27}{4}{43603}
\doinumber{10.5488/CMP.27.43603}

\title[Ising model in the R\'{e}nyi statistics]
{Ising model in the R\'{e}nyi statistics: the finite size effects}
\author[V. V.~Ignatyuk, A. P.~Moina]
{V. V.~Ignatyuk\orcid{0000-0003-4021-441X}\refaddr{label1},\refaddr{label2}\thanks{Corresponding author: \email{ignat@icmp.lviv.ua}.}
      \quad A. P.~Moina\orcid{0000-0002-2170-2445}\refaddr{label1}}

\addresses{
\addr{label1} Institute for Condensed Matter Physics, 1 Svientsitskii St., 79011 Lviv, Ukraine
\addr{label2} Ivan Franko National University of Lviv, 1 Universytetska St., Lviv, 79007, Ukraine}

\Keywords{R\'{e}nyi statistics, microcanocical ensemble, entropic phase transition, Ising model}

\newcommand{\beq}{\begin{equation}}
\newcommand{\eeq}{\end{equation}}
\newcommand{\bi}{\begin{itemize}}
	\newcommand{\ei}{\end{itemize}}
\newcommand{\bea}{\begin{eqnarray}}
\newcommand{\eea}{\end{eqnarray}}
\newcommand{\ban}{\begin{eqnarray*}}
	\newcommand{\ean}{\end{eqnarray*}}
\newcommand{\barr}{\begin{array}}
	\newcommand{\earr}{\end{array}}

\renewcommand{\theequation}{\arabic{section}.\arabic{equation}}

\date{Received September 26, 2024, in final form October 31, 2024}
\begin{document}

\maketitle

\begin{abstract}
The R\'{e}nyi statistics is applied for a description of finite size effects in the 1D Ising model. We calculate the internal energy of the spin chain and the system temperature using the R\'{e}nyi distribution and postulate them to be equal to their counterparts, obtained in the microcanonical ensemble. It allows us to self-consistently derive the R\'{e}nyi $q$-index and the Lagrange parameter $T$ to relate them to the physically observed system temperature $T_{\rm ph}$, and to show that the entropic phase transitions are possible in a broad temperature domain. We have also studied the temperature dependence of the internal energy $U(T_{\rm ph})$ at constant $q$ and an influence of the size related effects on the system thermodynamics.

\printkeywords
%
\end{abstract}

\section{Introduction}

By the early 20-th century, the basic principles of statistical description of thermodynamic
systems had been well established by the efforts of Boltzmann, Gibbs, Einstein, and many others. In 1957, Jaynes~\cite{Jaynes57}
formulated the probabilistic approach based on the knowledge of some statistical information about the system and introduced the rule, now widely known as the maximum entropy principle (MEP); when applied to the statistical mechanics it yields the famous Gibbs canonical distribution. Today this rule is used in many fields of science, ranging from physics and chemistry to the stock market analysis.

Further advances were stipulated by the growing necessity to generalize the elaborated schemes onto description of the systems, consisting of large but finite numbers of particles, for which the thermodynamic limit cannot be performed. The recent achievements in the study of a microcosm rely not only on more sophisticated experimental measurements \cite{single1,single2}, but also on a creation of the accompanying theoretical methods such as the nano-thermodynamics \cite{nanoTD1,nanoTD2}.

The second way to proceed is a construction of the groundwork of the non-equilibrium statistical mechanics and thermodynamics \cite{ZMR}. Obviously, a description of the non-equilibrium properties of
small sized objects becomes much more complicated, due to: i) large fluctuations in the system, ii) a presence of the local equilibrium or steady states rather than a true thermodynamic equilibrium and iii) a necessity to use the non-Gibbsian statistics when describing the observables. Nevertheless, serious advances have been recently made in a development of the theory of open quantum systems at the finite sizes of their environments \cite{open-micro2007,open-micro2022}, when the microcanonical distribution was used to derive the corresponding master equations and to calculate the non-equilibrium mean values.

Conceptually different methods of handling the finite size systems rest upon the usage of other non-Gibbsian statistics. The basics of the so-called parastatistics were introduced in the pioneering papers by R\'{e}nyi \cite{RenyiSP} and Tsallis \cite{Tsallis88,Tsallis-Ent} and were developed by many others, e.g., \cite{Sharma,Abe,Bashkirov-Book}.
At the first stage, the entropy functional is constructed within the framework of one or another parastatistics. A common feature of all the cases is an application of the MEP under a demand for the internal energy to be fixed and the availability of the normalization
condition. This conditional extremum problem yields  explicit expressions for the distribution functions.

Though an applications of the R\'{e}nyi and Tsallis statistics are wide enough, ranging from the description
of the systems of non-physical nature (Zipf's law) up to the thermodynamics of black holes~\cite{Bashkirov-Book}, the possibility to use these distributions to study the finite size objects is very important. Here, we mean the cases of finite environments regardless of how large are the systems interacting with them provided these systems are finite as well.
In this context, the above mentioned statistics can be viewed as a certain alternative for the microcanonical distribution for investigation of the systems that do not allow the thermodynamic limit to be performed, despite the fact that these statistics are based on completely different foundations. The R\'{e}nyi statistics seem to be most promising, being the additive one, unlike the Thallis statistics that are inherently non-additive and cannot be linked in any way to the microcanonical ensemble, which also completely rests upon the additivity principle.

Application of the R\'{e}nyi statistics opens up new prospects and sheds light on the specific features that are not met when we use the microcanonical distribution. Firstly, we mean a possibility for the entropic phase transitions to appear, which are not observed when we use the traditional (canonical or microcanonical) ensembles. In this paper, we put a question whether the inherently different R\'{e}nyi and microcanonical distributions could render a similar thermodynamics. In other words, if one postulates that the observables obtained within the two statistics are the same or very close, then how and to what extent the non-observable characteristics (for instance, the system entropy) will differ. We study the behavior of change of the R\'{e}nyi index $q$ at the entropic phase transition and compare it with the results of other theories, where it is determined from the heat capacity of the environment.

The paper is structured as follows. In section~\ref{sec2}, we present the basic relations of the R\'{e}nyi thermostatistics, which follow from the MEP. The limiting cases of $q\to 0$ and $q\to 1$, leading to the microcanonical distribution and the Gibbs statistics, are emphasized. In section~\ref{sec3}, we discuss some remarkable features typical of the R\'{e}nyi statistics: i) the entropic phase transition, which appears at $q\to 1$, and ii) the relation between the R\'{e}nyi index $q$ and the specific heat of the system environment, which comes after an application of the thermal fluctuations approach \cite{Bashkirov-Book} or more sophisticated exact method \cite{Almeida}. In section~\ref{sec4}, we consider a chain of Ising spins and present some expressions describing its equilibrium properties in the microcanonical ensemble.

The most substantial is section~\ref{sec5}, where the Ising model is studied within the R\'{e}nyi statistics. First, having assumed that the observables (internal energies and temperatures) are equal in the microcanonical and R\'{e}nyi ensembles, we show that the entropic phase transition appears in the second case and that all the remarkable relations mentioned in section~\ref{sec3} are obeyed with a great accuracy. A special attention is paid to the description of the system on the microscopic level: we analyse the ``allowed'' and ``forbidden'' microstates, their origin and influence on the system behaviour via the corresponding macrostates formation. Second, we study the temperature dependence of the internal energy $U(T)$ at constant $q$ and the size related effects, by which we mean an influence of two factors: the discreteness and finiteness of the system.

In the final section, we summarize the obtained results and point out some open questions.

\section{The R\'{e}nyi entropy and the R\'{e}nyi distribution}\label{sec2}

We start in a somewhat formal way.
Using a notion of the ``free entropy'' \cite{Bashkirov-Book}, we can introduce the generating function
\bea\label{PhiS}
\Phi_S(q)=\sum\limits_{i=1}^W \re^{-q S_i^{(\text B)}},
\eea
where $S_i^{(\text B)}=-\ln p_i$ denotes the Boltzmann entropy of the ensemble of microcanonical subsystems with probabilities $p_i$, which are generally dependent on the state $i$. {Hereafter, we put the Boltzmann constant $k_{\text B}$ to be equal to the unity.}
The summation in (\ref{PhiS}) runs over all the energy states starting from the lowest one up to the value $W$ (a limiting case $W\to\infty$ can also be considered).
The cumulant generating function that corresponds to (\ref{PhiS}) looks as follows:
\bea\label{PsiS}
\Psi_S(q)=\ln\Phi_S(q)=\ln\sum\limits_{i=1}^W p_i^{q}.
\eea
Dividing it by $1-q$, we  obtain the sought ``free entropy'', which coincides with the known R\'{e}nyi entropy
\bea\label{SR}
S_q^{(\text{R})}(p)=\frac{1}{1-q}\ln\sum\limits_{i=1}^W p_i^{q},
\eea
and includes the Gibbs-Shannon entropy  $S^{(\text G)}=-\sum_{i=1}^W p_i\ln p_i$ as a particular case when $q\to 1$.

Before the explicit expressions for the probabilities $p_i$ are obtained, it is worth explaining the reason to introduce the family of entropies (\ref{SR}). First of all, the Gibbs-Shannon entropy derived by a simple averaging of the Boltzmann entropy cannot be the function, whose extremum characterizes the steady state of a complex system under the entropy exchange with the
surroundings \cite{Bashkirov-Book}. Therefore, it is pertinent to introduce a notion of the \textit{entropy bath} for such processes, just like  the concept of the \textit{thermal bath} has been adopted for a small subsystem being in contact with its surroundings, which is considered to be infinite in the thermodynamic sense. Coupling with the entropy bath is a necessary condition for the self-organization of a complex system \cite{Klimontovich}. As a result of such coupling, the system under consideration cannot reach the state of thermodynamic equilibrium characterized by the minimum of the Helmholtz free energy. 

Since the entropy flux is usually accompanied by the heat flux, it is almost impossible to exclude one of the above mentioned reservoirs from consideration and to focus only on the other one. That is why the concept of entropy bath is not very popular in the scientific literature. In our case, it serves mainly to derive the expressions for the R\'{e}nyi entropy in the most transparent way. Nevertheless, one can mention the so-called dephasing model \cite{PRA2012,PRA2015}, widely used in the theory of open quantum systems to describe the processes of pure decoherence/recoherence. In such a model, there is no energy exchange between the spin or spin chain and the bath; the populations of levels remain constant, but the off-diagonal elements of the density matrix related to the system coherence evolve in time. This leads to a decrease (or increase) of the Bloch vector modulus and, as a consequence, to the entropy flow from the system to its environment or vice versa. Though the dephasing model is exactly solvable at averaging over the Gibbs equilibrium distribution, some interesting results were recently obtained for a similar spin-boson system using the R\'{e}nyi ensemble \cite{Semin2020}. From our point of view, the application of the R\'{e}nyi statistics would be most promising for open quantum systems interacting with finite size environments. In the subsequent sections, we use the R\'{e}nyi distribution as a certain alternative to the microcanonical one when describing the thermodynamics of the Ising system with a finite number of spins.

Though the limit $q\to 1$ in the R\'{e}nyi entropy (\ref{SR}) resulting in its Gibbs-Shannon counterpart is widely applied, another limiting case $q\to 0$ is considered much more rarely \cite{Parvan1,Parvan2,Parvan3}. A simple inspection of equation~(\ref{SR}) shows that in this limit the R\'{e}nyi entropy converts into the system entropy determined in the microcanonical ensemble:
\bea\label{SRqto0}
S^{\rm micro}=\lim\limits_{q\to 0} S^{(\text R)}_q(p)=\ln W.
\eea
The microcanonical distribution is known to describe a behaviour of an adiabatically isolated system~\cite{Landau}, consisting of $N$ particles contained within the volume $V$. Correspondingly, its small subsystem is distributed canonically if the thermodynamic limit $N\to\infty$, $V\to\infty$ can be applied at the constant value $n=N/V$ of the number density. Our task is to combine these two limits: $q\to 1$ that corresponds to the infinite bath and $q\to 0$ that is realized if the surrounding of the subsystem is vanishing --- into a single case defined by the R\'{e}nyi distribution. 

To derive the expression for the R\'{e}nyi distribution, let us proceed in a standard manner, considering the conditional extremum of the entropy (\ref{SR}) under the corresponding constraints of the normalization condition and
constant internal energy:
\bea\label{extremum1}
L_\text{R}(p)=\frac{1}{1-q}\ln\sum\limits_{i=1}^W p_i^q-\alpha \sum\limits_{i=1}^W p_i-\beta \sum\limits_{i=1}^W H_i p_i.
\eea
Equating the variation of the functional (\ref{extremum1})  to zero,
\bea\label{extremum2}
\frac{\delta L_\text{R}(p)}{\delta p_i}=\frac{q}{1-q}\frac{p_i^{q-1}}{\sum_{i=1}^W {p_i^q}}-\alpha-\beta H_i=0,
\eea
we obtain the explicit expression for the R\'{e}nyi distribution,
\bea\label{pR}
p_i^{(\text{R})}\equiv p_i=\frac{1}{Z^{(\text{R})}_q}\left(1-\beta \frac{q-1}{q}\Delta H_i\right)^{1/(q-1)},
\eea
where
\bea\label{ZR}
Z^{(\text{R})}_q=\sum\limits_{i=1}^W\left(1-\beta \frac{q-1}{q}\Delta H_i\right)^{1/(q-1)}
\eea
denotes the partition function, while the deviation of the system energy from its mean value $\Delta H_i\equiv H_i-U$ should be obtained from the self-consistency condition for the internal energy $U$,
\bea\label{selfU}
U=\sum\limits_{i=1}^W H_i p_i.
\eea
{As in the case of the Gibbs statistics, the so far unspecified Lagrange parameter $\beta$ can be associated with the inverse temperature, $\beta=T^{-1}$, expressed in the energy units since $k_{\text B}\equiv 1$;
further it will be convenient to position alternatively the temperature $T$ as the Lagrange multiplier as well.}
	Unlike the Gibbs case, the internal energy $U$ both enters the R\'{e}nyi distribution and is defined by the $p_i$ itself, as it follows from~(\ref{pR})--(\ref{selfU}). At the canonical distribution, which is obtained when the power law function~(\ref{pR}) converts into the exponential form at $q\to 1$, the internal energy $U$ enters both the numerator and denominator (the corresponding partition function $Z^{(\text{G})}$) and is mutually reduced.

There is an alternative representation for the R\'{e}nyi distribution, which can be obtained when the constraint (\ref{selfU}) is changed to the averaging over the so-called escort distribution $p_i^{\rm esc}$\cite{Bashkirov-Book,Lenzi2000}:
\bea\label{self-Uesc}
U=\sum\limits_{i=1}^W H_i p_i^{\rm esc},\qquad p_i^{\rm esc}=\frac{p_i^q}{\sum_{i=1}^W p_i^q}.
\eea
The corresponding R\'{e}nyi distribution function $p_i^{\rm esc}$ can be written down as follows:
\bea\label{pResc}
p_i^{\rm esc}=\frac{1}{Z^{\rm esc}}\left[1-\beta (1-q)\Delta H_i\right]^{1/(1-q)},\qquad Z^{\rm esc}=\sum\limits_{i=1}^W \left[1-\beta (1-q)\Delta H_i\right]^{1/(1-q)}.
\eea
The detailed explanation of the differences between the distributions (\ref{pR}) and (\ref{pResc}) and the physical arguments to choose one or the other can be found, e.g., in \cite{Lenzi2000}. Since in this paper we opted for the distribution (\ref{pR}), we are not going to explore this subject further, only mentioning that this concept is very close to the arguments\footnote{The expression for the non-extensive Tsallis entropy can be formally obtained after series expansion of (\ref{pR}) at $q=1$, $S^{(T)}_q(p)=\frac{1}{1-q}\left(\sum_{i=1}^W p_i^q-1
	\right)$, which is equivalent to the condition $|1-q|\ll 1$.} 
in favour of the 3-rd form of the Tsallis statistics over the 1-st and the 2-nd forms \cite{Tsallis1998}.

\section{Some remarkable relations typical of the R\'{e}nyi distribution}\label{sec3}

In this section, we are going to present some known relations dealt with the R\'{e}nyi distribution, which we shall extensively use, when analyzing the data of numerical calculations. It is straightforward to show that the R\'{e}nyi entropy (\ref{SR}) can be expressed via the partition function (\ref{ZR}) in a way similar to the microcanonical ensemble:
\bea\label{SR2}
S^{(\text{R})}_{\eta}= \ln Z^{(\text{R})}_{\eta}= \ln\sum\limits_{i=1}^W \left(1+\beta\frac{\eta}{\eta-1}\Delta H_i
\right)^{-1/\eta},
\eea
where the new constant $\eta=1-q$ is introduced for some physical reasons to be explained soon. When $\eta\to 0$, this entropy turns into the Gibbs entropy 
\bea\label{SG}
S^{(\text G)}=\ln\sum\limits_{i=1}^W \re^{-\beta\Delta H_i}.
\eea
Taking the derivative of the entropies difference $\Delta S=S^{(\text{R})}_{\eta}-S^{(\text G)}$ with respect to $\eta$, one can easily obtain the following relation:
\bea\label{relat1}
\lim\limits_{\eta\to 0}\frac{\rd \Delta S}{\rd\eta}=\frac{\beta^2}{2}\sum\limits_{i=1}^W p_i^{(\text G)}(\Delta H_i)^2,
\eea
where 
$\displaystyle p_i^{(\text G)}=\frac{\re^{-\beta H_i}}{\sum_{i=1}^W \re^{-\beta H_i}}$
denotes the Gibbs distribution. The expression in the r.h.s. of equation~(\ref{relat1}) can be rewritten in a more convenient way using a simple thermodynamic relation, 
\bea\label{rhs1}
\sum\limits_{i=1}^W p_i^{(\text G)}(\Delta H_i)^2=\frac{1}{\beta^2}\frac{\rd U}{\rd T}=\frac{1}{\beta^2} C_V,
\eea
where $U$ means the internal energy (\ref{selfU}), and $C_V$ denotes the heat capacity at constant volume.

Combining equations~(\ref{relat1}) and (\ref{rhs1}), we arrive at {the first} remarkable relation:
\bea\label{relat1fin}
\lim\limits_{\eta\to 0}\frac{\rd \Delta S}{\rd \eta}=\frac{1}{2}C_V.
\eea
In spite of its simple form, this expression has a profound physical meaning. If we relate $\eta$ to some order parameter, it enables us to consider the transition to the R\'{e}nyi thermostatistics as a peculiar kind of the phase transition
into a more organized state. Following \cite{Bashkirov-Book}, we call it the \textit{entropic phase transition}.
As a result of the entropic phase transition, the system passes into an
ordered state with the order parameter $\eta\ne 0$. In contrast to the usual phase transition that takes place at some critical temperature $T_{\rm cr}$, which can be evaluated taking the thermodynamic limit, conditions of the entropic phase transition will be shown to depend on the size of the finite system.

Note that relation (\ref{relat1fin}) is exact, since no assumption or approximation has been made at its derivation. The physical reasoning, leading to {the second} useful relation, equation (\ref{relat3}), is semi-phenomenological and less rigorous. Nevertheless, the said relation is just  as important as equation~(\ref{relat1fin}). Since its derivation needs some intermediate calculations, which are not relevant in our case, we shall only briefly outline the basic points underlying its origin (see \cite{Bashkirov-Book} and references therein).

Suppose that the finite system (hereafter called the subsystem) is placed in the environment of finite size and heat capacity. This environment defines the temperature $T$ of the total combined system, which can fluctuate around some average value $T_0$. 
The amount of heat transferred from the environment to the subsystem  is finite, and the time of temperature equilibration distinctly differs from zero causing the mentioned fluctuations. The time behaviour of temperature can be described by the Langevin equation with white noise. At the next stage of description, one can pass to the corresponding Fokker-Planck equation for the distribution function of random temperature $T$, which stationary solution can be obtained explicitly. A subsequent assumption is very similar to that underlying the concept of local equilibrium, which is widely used in the problems of non-equilibrium statistical mechanics \cite{ZMR}. Let us assume that any mesoscopical part of the total system obeys the canonical distribution, but the temperatures of these small subsystems are random values. With the stationary solution $f(T)$ of the Fokker-Planck equation in hand, and averaging the Gibbs distributions $p^{(\text G)}(T)$ over $f(T)$, one can obtain the final distribution function governing a thermostatistics of the above system. The obtained formula can be related to the R\'{e}nyi distribution function, if one identifies 
\bea\label{relat3}
q=1-\frac{1}{C_{VE}}
\eea
and
\bea\label{relat4}
T_0=q T,
\eea
where $C_{VE}$ is the heat capacity of the environment.

The above presented approach is quite illustrative in some physical sense, but definitely lacks rigo\-rous\-ness; it is obvious that there are  lots possibilities \cite{PhysA2019} to obtain the probability density $f(T)$. However, more rigorous approaches (see, for instance, \cite{Almeida}) are based on the exact relations of statistical mechanics and yield the results similar to (\ref{relat3}).

The relations (\ref{relat3})--(\ref{relat4}) are worthy to be commented a bit more. First of all, the order parameter $\eta=(1-q)$ is inversely proportional to the number of particles in the environment $N-N_L$, since the heat capacity of the environment $C_{VE}\sim (N-N_L)$. Note that the subsystem consists of $N_L$ particles, and its heat capacity in equation~(\ref{relat1fin}) is proportional to $N_L$. In this context, the order parameter $\eta$, being neither purely intensive nor extensive, differs from those known in the phase transition theory. Obviously, in the thermodynamic limit $N\to\infty$, $V\to\infty$, $N/V=n=\mbox{const}$, and at $N_L=\mbox{const}$, the $\eta$ tends to zero, yielding the canonical distribution.

The second point follows directly from equation~(\ref{relat4}). The meaning of physically measurable temperature can be attributed to $T_0$, as it corresponds to the average value of the system temperature, rather than to the fluctuating temperature $T$ dealt with the Lagrange multiplier $\beta=1/T$, see equation~(\ref{extremum1}). In the limit $\eta\to 0$, the temperature $T$ becomes $\delta$-distributed around $T_0$, $f(T)=\delta(T-T_0)$, and the two temperatures coincide, $T_0=T$.

The final point to be mentioned in this section is that the relation (\ref{relat3}) changes to 
\bea\label{relatEsc}
q=1+\frac{1}{C_{VE}},
\eea
if one uses the R\'{e}nyi distribution (\ref{pResc}) instead of (\ref{pR}).
Equation~(\ref{relatEsc}) assumes that the R\'{e}nyi index can be greater than unity, in contrast to the case (\ref{relat3}). In section~\ref{sec5}, we will use equations~(\ref{relat1fin})--(\ref{relatEsc}) as the reference relations to verify the consistency of the proposed approach.

\section{The Ising model in the microcanonical ensemble}\label{sec4}

Let us consider an isolated one-dimensional chain of $N$ Ising spins with periodic boundary conditions. The Hamiltonian of such a system can be written down in a usual form:
\bea\label{H}
H=-J\sum\limits_{i=1}^N\sigma_i\sigma_{i+1},
\eea
where $J$ denotes the interaction strength, and $\sigma_i=\pm 1$. Possible values of the system energy are determined by an even integer $M$ ($0 \leqslant M\leqslant N$): the  number of pairs\footnote{Hereafter, we follow the reasoning and use denotations for the corresponding values as those adopted in \cite{MSU}.}
of the oppositely directed neighbouring spins, $\sigma_i=-\sigma_{i+1}$.
The energy of such a spin configuration is 
\bea\label{EM}
E(M)=-J(N-2M),
\eea
while the number of configurations with the energy $E(M)$ is
\bea\label{GammaM}
\Gamma(M)=2 \frac{N!}{M!(N-M)!},
\eea
and all theses states with the given energy $E(M)$ are equally probable.

If the coupling is ferromagnetic, $J>0$, then in the ground state all spins are aligned in the same direction, and $M=0$. On the other hand, if the coupling is antiferromagnetic,
$J<0$, in the ground state we have an alternating spin alignment, and $M=N$ for even $N$.
Excited energy levels correspond to configurations with $M > 0$ and $M<N$, respectively. Obviously, the numbers $M$ or ($N-M$) are directly related to the system temperature: the larger is the temperature, the more spins are flipped as compared to the ground state configuration due to thermal fluctuations.

Let us now single out a subsystem containing $L$ spins, starting from the first spin and up to the $L$-th one. Hereafter, we will call these $L$ spins as the  \textit{``subsystem''}, whereas the spins from the $(L+1)$-th up to the $N$-th constitute the \textit{``bath''} or the \textit{``environment''}. There is an energy exchange between the subsystem and the bath since the 
``boundary'' spins interact with each other: the 1-st spin is coupled to the $N$-th one, and the $L$-th spin --- to the $(L+1)$-th one. The corresponding energy of the subsystem of $L$ spins is determined by the integer number $K$ of pairs of the oppositely directed neighbouring spins in it.\footnote{Since the chosen subsystem is an open spin chain, the values of $K$ can be either even or odd, in contrast to $M$, which must be even due to the periodic conditions imposed to the entire chain.}
This energy can be presented in a way similar to (\ref{EM}):
\bea\label{EK}
E_S(K)=-J(L-1-2 K),
\eea
while the number of the subsystem configurations with such an energy is
\bea\label{GammaK}
\Gamma_S(K)=2 \frac{(L-1)!}{K!(L-1-K)!}.
\eea
Within our consideration, the integer $K$ has the physical meaning of the possible number of the subsystem microstates and
obeys the following inequality:
\bea\label{KMinMax}
&&
K_{\rm min}\leqslant K\leqslant K_{\rm max},\\
\nonumber
&&
K_{\rm min}=\mbox{max}(0, M-N+L-1),\qquad K_{\rm max}=\mbox{min}(M,L-1).
\eea

Since we assume the entire chain of $N$ spins with the fixed energy to be distributed microcanonically, the probability to find the subsystem of $L$ spins in the microstate $K$ is equal to
\bea\label{wK}
\omega(K)=\frac{1}{\Gamma(M)}\frac{(N-L+1)!}{(M-K)!(N-L+1-M+K)!}.
\eea
The binomial coefficient in the r.h.s. of (\ref{wK}) determines the number of ways to choose from the $N-L+1$ spins, which do not belong to the subsystem, exactly  $M-K$ pairs of oppositely directed neighbouring spins, while the division by the total number of states (\ref{GammaM}) appears naturally due to the microcanonical distribution.

Using equations~(\ref{EK})--(\ref{wK}) and the properties of the Vandermonde's convolution \cite{MSU}, it is straightforward to calculate the mean energy of the subsystem as
\bea\label{ESmean}
\langle E_S\rangle\equiv\sum\limits_{K=K_{\rm min}}^{K_{\rm max}}\omega(K) \Gamma_S(K) E_S(K)=-J (L-1)\left(1-2\frac{M}{N}
\right).
\eea
Now let us turn our attention to transition from the microcanonical ensemble to the canonical one. It is known to occur in the thermodynamic limit, when the subsystem is infinitesimally small as compared to the bath: $N\to\infty, M\to\infty$, while $L$ and the ratio $a=M/N$ remain finite. In such a case, due to (\ref{KMinMax}) the number of subsystem microstates is determined by the inequality $0\leqslant K\leqslant L-1$. Besides, the following strong inequalities are also valid:
\bea\label{strong}
L-1\ll N,\qquad K\ll M, \qquad L-1-K\ll N-M.
\eea
Thus, using the approximation $(N-n)!\approx N!/N^n$, which is valid at $n\ll N$, one can pass from the probability (\ref{wK}) to the canonical Gibbs distribution $\omega^{(\text{G})}(K)=Z^{-1} \exp[-E(K)/\Theta]$, where the temperature $\Theta$ is equal to 
\bea\label{Theta}
\Theta=-2 J\Big/\ln \left(\frac{M}{N-M}
\right)=-2 J\Big/\ln \left(\frac{a}{1-a}
\right),
\eea
and the partition function $Z$ is expressed as
\bea\label{ZGibbs}
Z=2 \left[{a(1-a)}
\right]^{-\frac{L-1}{2}}.
\eea
As expected, the partition function depends on the subsystem size $L$ and temperature $\Theta$ and does not depend on the size of the thermal bath. The second point is that in the ferromagnetic case $(J>0)$ at $a>1/2$, the temperature (\ref{Theta}) becomes formally negative. The same is true in the antiferromagnetic case $(J<0)$ at $a<1/2$. A notion of the negative temperature is perfectly analysed in the well-known textbook \cite{Landau}, and we are not going to explore this subject any more. We just would like to note that for the ferromagnetic coupling $J>0$, the configuration, when the upside and downside oriented spins strictly alternate and $M=N$, corresponds to the temperature $\Theta\to -0$ and to the maximum of energy $E=J N$, whereas the case $M=N/2$ corresponds to $\Theta\to -\infty$ and the zero energy of the total system.

The above expressions are used in the next section, where we attempt to relate the observables, obtained within the microcanonical ensemble, to those in the R\'{e}nyi thermostatistics. The key point is that both ensembles allow one to take into consideration the effects when the size of a subsystem is comparable to that of its surroundings, or even becomes larger than this one.

\section{The Ising model in the R\'{e}nyi ensemble}\label{sec5}

Now let us combine and generalize the results presented in sections~\ref{sec2} and \ref{sec4} to study the behaviour of a finite Ising chain in the R\'{e}nyi statistics. First of all, using equations~(\ref{pR}) and (\ref{EK})--(\ref{GammaK}), we write down the explicit form of the R\'{e}nyi distribution for the chain of $L$ spins:
\bea\label{pRIsing}
p_\text{R}(L,K,q,T)=\Gamma_S(K)\left\{1-\frac{q-1}{q T}\left[-J(L-1-2 K)-U\right]
\right\}^{1/(q-1)}\Big/Z_\text{R}(L,q,T;U),
\eea
where the corresponding partition function reads as
\bea\label{ZRIsing}
Z_\text{R}(L,q,T;U)=\sum\limits_{K=K_{\rm min}}^{K_{\rm max}}\Gamma_S(K)\left\{1-\frac{q-1}{q T}\left[-J(L-1-2 K)-U\right]
\right\}^{1/(q-1)},
\eea
and, according to equation~(\ref{SR2}), defines the R\'{e}nyi entropy
\bea\label{SRIsing}
S_\text{R}(L,q,T;U)= \ln Z_\text{R}(L,q,T;U).
\eea
The self-consistency condition is expressed in the usual way, see equation~(\ref{selfU}):
\bea\label{selfUIsing}
U\equiv U(L,q,T)=\sum\limits_{K=K_{\rm min}}^{K_{\rm max}}p_\text{R}(L,K,q,T)E_S(K).
\eea

An essential point must be mentioned here. Since the R\'{e}nyi distribution (\ref{pRIsing}) is an exponential function with (in general) arbitrary power index $1/(q-1)$, the expressions in parentheses should be non-negative \cite{Lenzi2000}.  One can consider this condition as a strict requirement for the microstates with some ``unphysical''  values of $K$ to be forbidden. Consequently, there appear two possible ways to manage this problem in calculations: 
\begin{itemize}\label{2conditions}
	\item 
	\textbf{Option 1}: to  reject the entire macrostate if at least for one of its microstates the expression in the parentheses is negative;
	\item
	\textbf{Option 2}: to reject only the microstates with  ``forbidden'' $K$, while retaining the corresponding macrostate as a whole.
	Technically, it means mutiplying each parantheses in (\ref{pRIsing})--(\ref{ZRIsing}) by the appropriate Heaviside $\theta$-function, $\{...\}^{1/(q-1)}\to \{...\}^{1/(q-1)}_+=\{...\}^{1/(q-1)}\theta(...)$.
\end{itemize}
\label{options}
By the term \textit{``macrostate''} we mean a state of the system in the R\'{e}nyi ensemble with given values of $q$ and $T$ at the fixed length $L$ of the spin chain. In the subsequent numerical calculations  we shall consider both options.

Before studying the thermodynamic properties of the Ising model in the R\'{e}nyi ensemble and comparing them with those in the microcanonical one, we have to define two temperatures, related to the above statistics. These temperatures are expressed quite similarly via the corresponding entropies. In particular [see also equation~(\ref{SR2}) for comparison], the system entropy in the microcanonical ensemble is determined as 
\bea\label{Smicro}
S_{\rm micro}=\ln \Gamma(M), 
\eea
where the statistical weight $\Gamma(M)$ is defined in equation~(\ref{GammaM}). To proceed, we express factorials in the binomial coefficient (\ref{GammaM}) via the elementary functions using the Stirling approximation
\bea\label{Stirling}
N!\approx N^N\exp(-N)\sqrt{2\piup N}.
\eea
Now the entropy formally becomes a differentiable function, and the system temperature in the microcanonical ensemble can be calculated in the usual way:
\bea\label{Tmicro}
T_{\rm micro}=\left(\frac{\partial S}{\partial E}\right)^{-1}=\frac{\partial E(M)}{\partial M}\Big/\frac{\partial S(M)}{\partial M}=\frac{4 J M(N-M)}{2M-N-2M(N-M)\ln[M/({N-M})]}.
\eea
In the thermodynamic limit, it transforms into the temperature $\Theta$ defined in the
canonical ensemble, see equation~(\ref{Theta}). In a general case, the temperature $T_{\rm micro}$ depends not only on the ratio $a=M/N$, but also on the total spin number $N$. It can be easily shown that the inverse temperature $\beta_{\rm micro}\equiv 1/T_{\rm micro}$ scales as $\beta_{\rm micro}=\Theta^{-1}(1+b/N)$, where $b$ is a certain parameter independent of $N$. Therefore, the temperature of a finite system, which is determined in the microcanonical ensemble, can be positioned as a pseudo-intensive quantity \cite{axioms}, as opposed to the temperature derived with the Gibbs distribution. We will explore this point further when discussing the numerical results for the system entropy in the R\'{e}nyi thermostatistics.

The subsystem temperature defined in the R\'{e}nyi ensemble can be obtained similarly to (\ref{Tmicro}):
\bea\label{TR}
&&
T^{(\text{R})}\equiv T_{\rm ph}=\left(\frac{\partial S_\text{R}}{\partial U}
\right)^{-1}=q T Z_\text{R}/\tilde{Z}_\text{R},\\
\nonumber
&&
\tilde{Z}_\text{R}=\sum\limits_{K=K_{\rm min}}^{K_{\rm max}}
\Gamma_S(K)\left\{1-\frac{q-1}{q T}\left[-J(L-1-2 K)-U\right]
\right\}^{-1+1/(q-1)}.
\eea
We have omitted the arguments of the entropy function (\ref{SRIsing}) in equation~(\ref{TR}) and used a denotation $T_{\rm ph}$ for the temperature defined in the R\'{e}nyi ensemble. This is in line with the concept used in \cite{preprint} to introduce the ``physically measurable'' temperature in different parastatistics.

It can be shown by the series expansion of (\ref{TR}) at $q=1$ that at small deviations $|q-1|\ll 1$, the R\'{e}nyi temperature is expressed via the second $\langle \Delta H^2\rangle$ and the third $\langle \Delta H^3\rangle$ moments of the energy fluctuation, which should be calculated in the canonical ensemble. The linear term in the expansion disappears due to the imposed self-consistency condition $\langle\Delta H\rangle\equiv 0$, c.f. equation~(\ref{selfU}). The second moment $\langle \Delta H^2\rangle$ is given by (\ref{rhs1}). In a similar way, one can derive the expression for the third moment $\langle \Delta H^3\rangle$ and to write down the approximate expression for the physical temperature as follows:

	\bea\label{TRapprox}
	T_{\rm ph}\approx q T\left\{ 
	1+(q-1)^2 \left[
	\frac{T}{2}\frac{\rd C_V^{(\text{G})}}{\rd T}-C_V^{(\text{G})}
	\right]	
	\right\}=
	{
	q T\left\{ 
	1+(q-1)^2(L-1)\frac{J\tanh\left(\frac JT\right)-2T}{T^3 \cosh^2\left(\frac JT\right)} 	\right\},
}
	\eea
where we used	the heat capacity of the subsystem $C_V^{(\text{G})}$, expressed \cite{MSU} via the canonlical internal energy.
	\bea\label{CvG}
	C_V^{(\text{G})}(T)=\frac{\rd U^{(\text{G})}(T)}{\rd T},\qquad U^{(\text{G})}(T)=-J(L-1)\tanh\left(\frac{J}{T}\right).
	\eea

As it follows from equation (\ref{TRapprox}), the corrections to $q T$  vanish at small deviations of the R\'{e}nyi index from the unity { and for small subsystem sizes $L$. Besides, the corrections approach zero as $T^{-1}$ at $T\to\infty$ and much faster, as $\exp\left(\frac{-2J}{T}\right)T^{-2}$, at $T\to 0$. Hence, at $|q-1|\ll1$ or at very low or sufficiently high $T$,} the physical temperature should be well approximated by the expression $T_{\rm ph}\approx q T $, see also equation~(\ref{relat4}) for comparison.
The R\'{e}nyi temperature (\ref{TR}) can be considered as a generalization of the widely used linear dependence $T_{\rm ph}= q T $ \cite{Bashkirov-Book} for large deviations of $q$-index from the unity that occur when the subsystem is comparable with or even larger than the environment.

Beside the length of the spin chain $L$ and the microstate number $K$, the R\'{e}nyi distribution (\ref{pRIsing}) depends also on the R\'{e}nyi index $q$ and the Lagrange parameter $T$. Usually, the value of $q$ is determined by the choice of the system environment. In particular, in \cite{preprint} a similar chain of the Ising spins studied within the Tsallis statistics is assumed to be coupled to a thermal bath of an ideal gas. The index $q$ has been related to the heat capacity of the bath by an expression like (\ref{relat3}). However, at the beginning, we proceed along another pathway, which appears to be more self-consistent and refined in some sense.

\subsection{Problem I}
As a basis of further investigation, we consider the following
\underline{\textbf{assumption:}}
\\[1ex]
\noindent\textit{At certain values of the parameter $T^*$ and R\'{e}nyi index $q^*$, the physical temperature and the internal energy coincide with their counterparts calculated in the microcanonical ensemble,}
\begin{subequations}
	\label{selfC-main}
	\begin{align}
&	T_{\rm ph}(L,q^*,T^*)=T_{\rm micro}(N,M),\\
&	U(L,q^*,T^*)=\left\langle E_S(N,M,L)\right\rangle.
	\end{align}
\end{subequations}

\label{assumption}
From all possible solutions of equations~(\ref{selfC-main}), we choose that pair ($q^*,T^*$), which provides the largest value of the entropy (\ref{SRIsing}). In fact, we have a double maximization over the entropy, since initially we solved the conditional extremum problem for the functional (\ref{extremum1}) to obtain the explicit expression for the R\'{e}nyi distribution at arbitrary $q$.

The conditions (\ref{selfC-main}) appear to be physically well justified. Indeed, equation~(\ref{selfC-main}a) resembles the condition (28) of \cite{Lenzi2000} imposed on the R\'{e}nyi temperature to ensure the proper system thermodynamics in the R\'{e}nyi statistics. The only difference is that, in our case, we relate the R\'{e}nyi temperature with its microcanonical counterpart (\ref{Tmicro}) rather than with the canonical one, as it was done in \cite{Lenzi2000}. 
The assumption identical to equation~(\ref{selfC-main}a) has been also made in \cite{japan1,japan2}, where the classical and quantum systems governed  by the Tsallis and R\'{e}nyi statistics were considered. 

As for equation~(\ref{selfC-main}b), the situation is more subtle. It was shown in  \cite{japan2} that internal energies of the classical independent harmonic oscillators in the microcanonical and R\'{e}nyi statistics are identical as long as the relation $T_{\rm ph}=T_{\rm micro}$ is valid. In the case of quantum non-interacting harmonic oscillators, the same is true at high temperatures. In the general case \cite{japan1}, there is a difference between the internal energies $U$ and $\langle E_S\rangle$, which was found to be small if $(q-1)N<1$. We shall return to this issue in the final part of section~\ref{sec5}. However, at this stage of investigation, we believe the second equation in (\ref{selfC-main}) to be quite reasonable too, since we are trying to find out whether the R\'{e}nyi statistics yield the same values of the thermodynamic observables as the microcanonical one. If the answer is yes, then we can use the R\'{e}nyi statistics to describe the system thermodynamics as a certain alternative to the microcanonical statistics because both of them use the similar concept of the system/environment finiteness.

From the mathematical point of view, the solutions to the system of equations (\ref{selfC-main}) give one the dependences $T^*(N,M,L)$ and $q^*(N,M,L)$. In other words, at fixed values of $N$, $M$ and $L$, the observables (\ref{selfC-main}) are identical in both ensembles as long as the corresponding R\'{e}nyi index and Lagrange parameter $T$ are equal to $q^*(N,M,L)$ and $T^*(N,M,L)$. However, the entropy calculated in the R\'{e}nyi and microcanonical (as well as canonical) ensembles will be different, see equations~(\ref{SRIsing}) and (\ref{Smicro}). To study  the entropy dependence on temperature and size of the subsystem is one of our main goals of this section.
\begin{figure}
	\centerline{\includegraphics[height=0.20\textheight, width=0.38\textwidth]{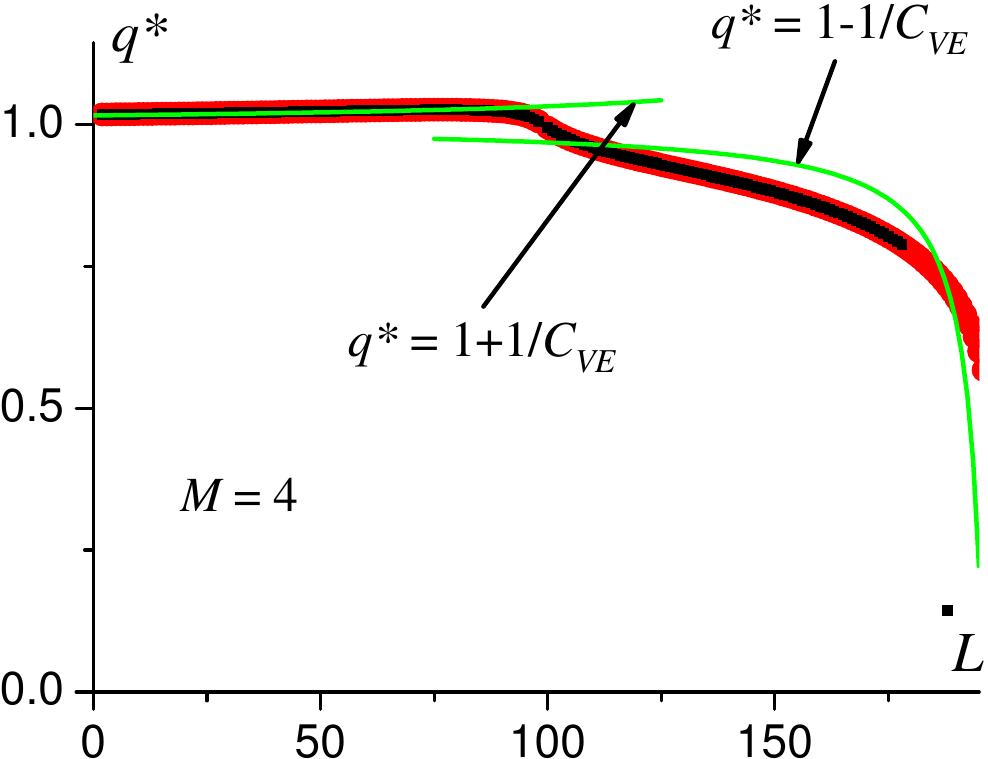}
		\hspace{5mm}
		\includegraphics[height=0.20\textheight, width=0.38\textwidth]{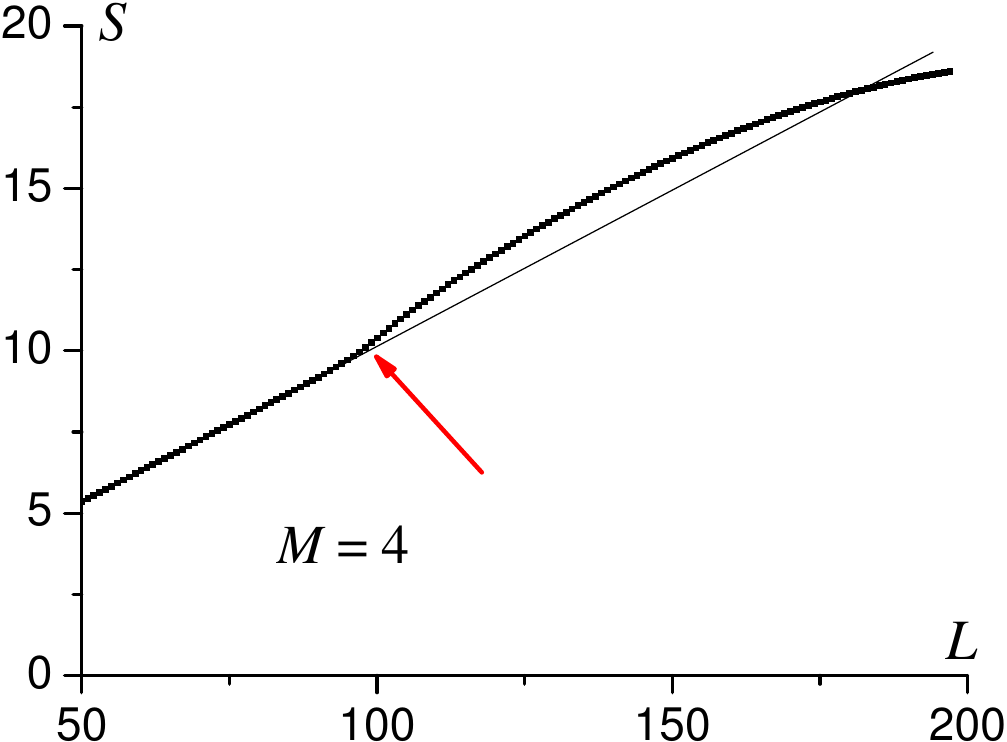}}
	\vspace{3mm}
	\centerline{\includegraphics[height=0.20\textheight, width=0.38\textwidth]{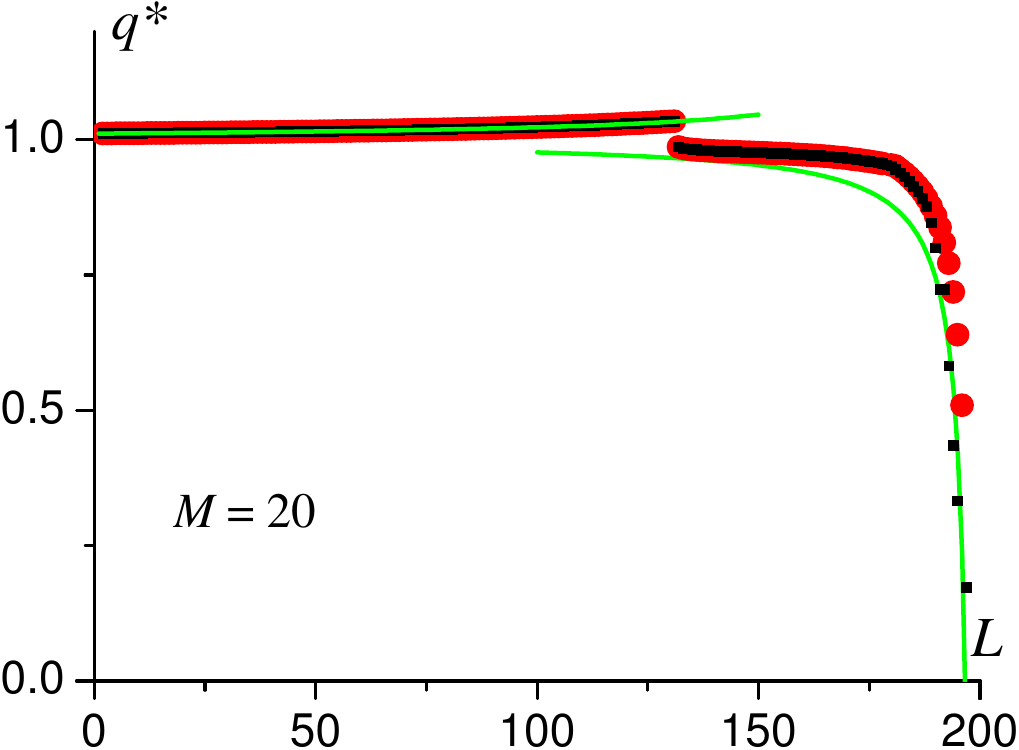}
		\hspace{5mm}
		\includegraphics[height=0.20\textheight, width=0.38\textwidth]{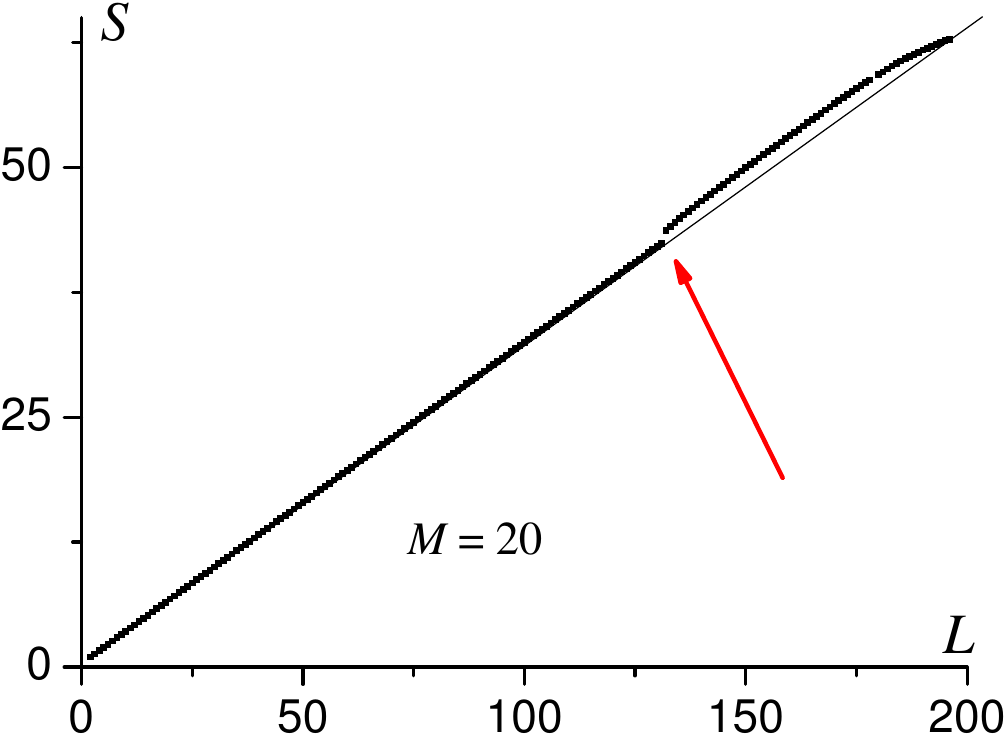}}
	\vspace{3mm}
	\centerline{\includegraphics[height=0.20\textheight, width=0.38\textwidth]{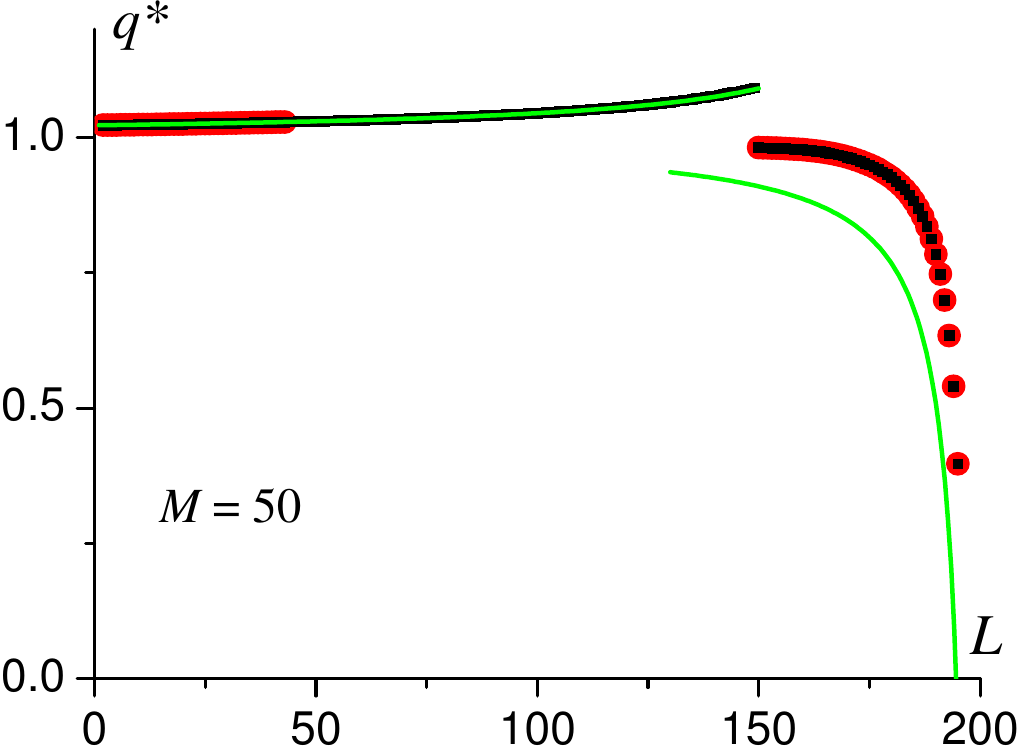}
		\hspace{5mm}
		\includegraphics[height=0.20\textheight, width=0.38\textwidth]{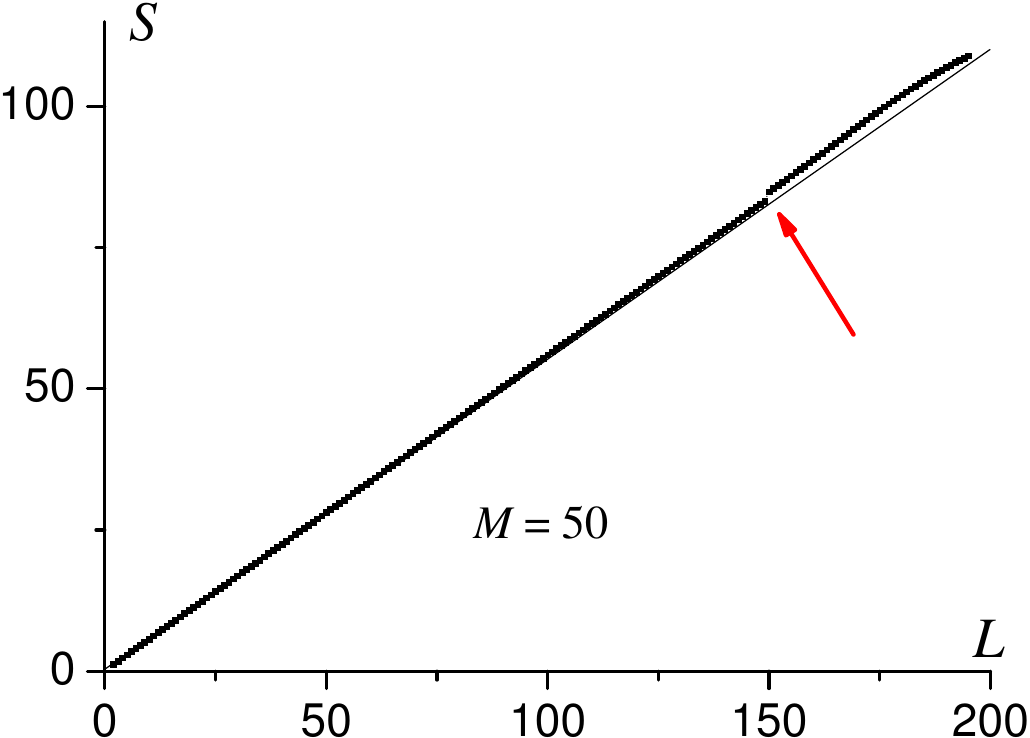}}
	\vspace{3mm}
	\centerline{\includegraphics[height=0.20\textheight, width=0.38\textwidth]{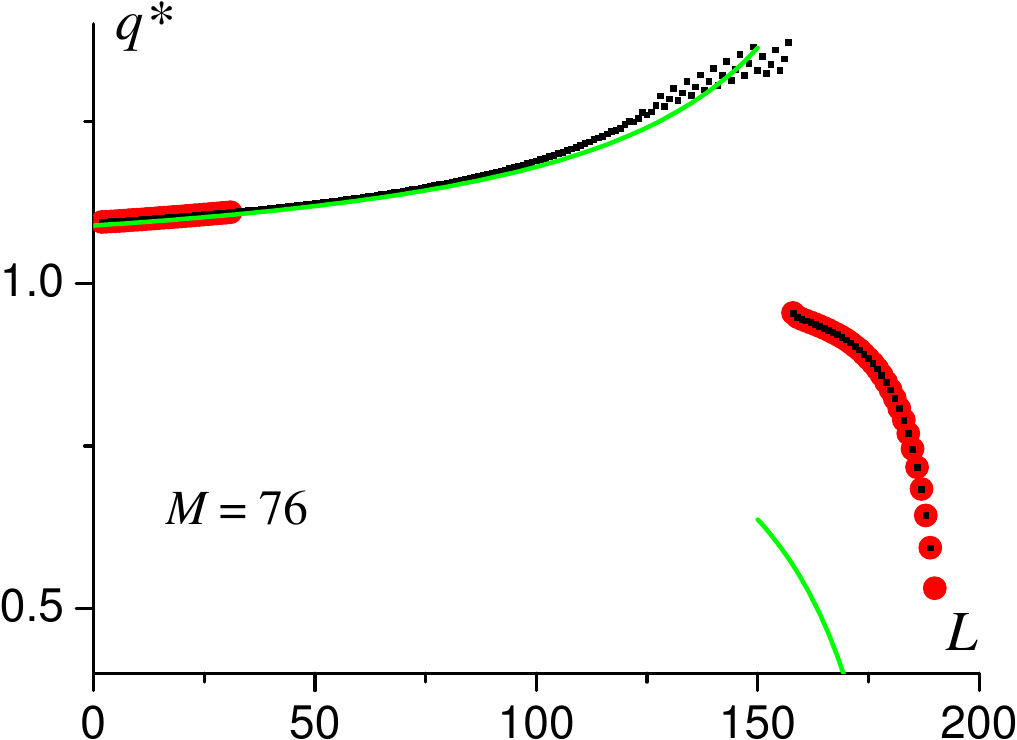}
		\hspace{5mm}
		\includegraphics[height=0.20\textheight, width=0.38\textwidth]{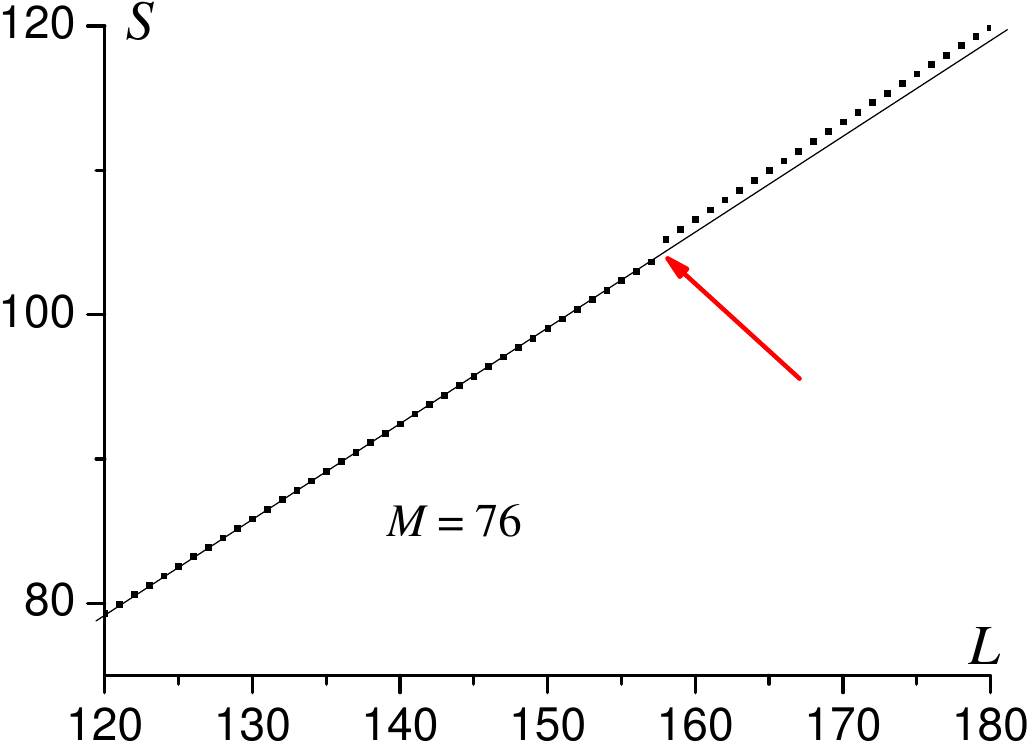}}
	\caption{(Colour online) Dependences of the R\'{e}nyi index $q^*$ (left) and the R\'{e}nyi entropy $S$ (right) on the subsystem size $L$ at different $M$ for $N=200$. The red circles and black squares are calculated using options 1 and 2, respectively (see page~\pageref{options}). The green lines are the $q^*= 1\pm 1/C_{VE}$ dependences found from equations~(\ref{relatEsc}) and (\ref{relat3}). Black lines are the extrapolations of the linear dependences $S(L)$ to large~$L$. The red arrows indicate the entropic phase transitions, where $\lim_{\eta\to 0}\partial S/\partial\eta= C_{V}/2$.}
	\label{qL-SL}
\end{figure}%

In figure \ref{qL-SL}, we plot the dependences of the R\'{e}nyi index $q^*$ and R\'{e}nyi entropy $S$ 
 on the number $L$ of spins in the subsystem at different values of $M$ for $N=200$. In all numerical calculations we take the ferromagnetic $J=1$. The function $q^*(L)$ is either continuous (at $M=2\div 6$ that corresponds to low temperatures) 
or discontinuous (at $M\geqslant 8$). { Obviously, the term ``continuity'' means a regular behaviour of the corresponding functions since the argument $L$ is itself discrete.}
In the former case (top row, $M=4$), something like the 2-nd order entropic phase transition occurs: at a certain value $L_{\rm trans}$, the entropy starts to deviate from the straight line, being at that a continuous function of the subsystem size $L$. Using the terminology adopted in the recent paper by Tsallis \cite{axioms}, the R\'{e}nyi entropy becomes a pseudo-extensive function of the subsystem size. 
There is a discrepancy between our results and the reasoning presented in the cited paper, which needs some explanation.
In \cite{axioms}, the particle number $N$, the system volume $V$, and the system entropy $S$ are imposed to be \textit{extensive} functions, whereas the thermodynamic potentials are allowed to be \textit{ pseudo-extensive}. In our case, the internal energy (\ref{ESmean}) is extensive by definition.
Hence, the entropy $S$ is forced to behave in a pseudo-extensive manner.\footnote{There is also a more formal explanation of such a behaviour of the entropy. Proceeding in a strict mathematical way and using two additional constraints (\ref{selfC-main}), one has to re-derive the distribution, which obviously could differ from the R\'{e}nyi probability~(\ref{pRIsing}). Since such a generalized problem of the conditional extremum seems to be unsolvable explicitly, we prefer to use the distribution~(\ref{pRIsing}) instead of the desired but unobtainable ``exact'' one. The pseudo-extensive behaviour of the entropy can be directly related to this fact.}.

Quite expectedly, at high $L\to N$  when $q\to 0$, the curve $S(L)$ approaches the linear extrapolation of its low $L$ part, and thus the microcanonical distribution of the Ising spins that ensures \textit{the extensive} behaviour of the system entropy is restored. Note that this is also true for other values of $M$, when a noticeable jump of $S(L)$ occurs at $L_{\rm trans}$, and we deal with
something like the 1-st order entropic phase transition, which is well-known to be absent in the canonical ensemble. We were pleased to note that despite the discreteness of our model and the impossibility to perform the continuous limit transition $q\to 1$ at the point of the above mentioned jump, a difference scheme approximation to the relation (\ref{relat1fin})
\bea\label{diff}
\lim\limits_{q\to 1}\left(\frac{\partial S}{\partial q}\right)=-\frac{1}{2}C_V\approx \frac{S(L_{\rm trans})-S(L_{\rm trans}+1)}{q(L_{\rm trans})-q(L_{\rm trans}+1)}
\eea
holds within the accuracy of a few percent (at least, at low to moderate $M$; see the discussion below). This once again indicates that the assumption (\ref{selfC-main}) is reasonable.
Here, $L_{\rm trans}$ is  the largest $L$ below the transition point, while $L_{\rm trans}+1$ is the smallest length above the transition point.
 The explicit expression for the heat capacity of the subsystem is obtained from a simple thermodynamic relation
\bea\label{CV}
C_V=\frac{\rd\langle E_S\rangle}{\rd T_{\rm micro}}=\frac{\rd\langle E_S\rangle}{\rd M}\Big/\frac{\rd T_{\rm micro}}{\rd M}=
-\frac{(L-1)[N-2M+2M(N-M)\ln(M/(N-M))]^2}{2N[N^2-2M(N-M)(N+1)]}
\eea
using the expressions (\ref{ESmean}) and (\ref{Tmicro}). The expression for the heat capacity $C_{VE}$ of the environment is found from (\ref{CV}) with a substitution $L\longrightarrow N-L$. Note that $C_{V}$ and $C_{VE}$ have maxima positioned at about $M/N=0.08$, which within the canonical distribution corresponds to temperature $\sim 0.82J$ [see equation~(\ref{Theta})].

It can be noted that the values of $q^*$ coincide with those predicted by (\ref{relatEsc}) below $L_{\rm trans}$, except for the domain of fast oscillations of $q^*$. In its turn, (\ref{relat3}) qualitatively reproduces the behaviour of $q^*(L)$  above $L_{\rm trans}$, with the best quantitative agreement  observed for $M$ close to the position of the maximum of $C_{VE}(M)$.
The differences are to be expected, since we used a completely different method of calculating $q$, not based on equations~(\ref{relatEsc}) or (\ref{relat3}). Even though expression (\ref{relatEsc}) arises due to the application of the escort distribution 
 (this can be a subject of a separate study), a direct comparison of our results   with the reference formulae mentioned in section~\ref{sec3} looks quite interesting.

The analysis of the results plotted in figure~\ref{qL-SL} brings us back to the ``\textbf{option 1} vs. \textbf{option 2}'' dilemma, see page~\pageref{options}. At low to moderate $M$ values, \textbf{option 1} (when the entire macrostate is rejected if at least one of the parentheses in the distribution (\ref{pRIsing}) is negative) appears much more favourable for the description of $q^{*}(L)$ of large subsystems, providing a smooth heading of the R\'{e}nyi index to zero. 

At the intermediate values of $M\sim 40\div 60$, the two options yield identical results for large $L$. However, below  $L_{\rm trans}$, there appears a gap in the  $q^*(L)$ dependence calculated according to \textbf{option 1}, while~\textbf{option~2} ensures the existence of the solutions of (\ref{selfC-main}) in the whole domain of $L$. This gap grows with $M$, and fast oscillations of $q^*(L)$, calculated according to \textbf{option 2}, emerge below $L_{\rm trans}$ (see the bottom left-hand part in figure~\ref{qL-SL}). We attribute this behaviour to the rapid changes in the number of ``unphysical'' microstates $K$  with $L$. It is closely related to the introduced notion of the size effects since we face a discreteness of $K$ and a finiteness of the subsystem size $L$. However, no such oscillation is observed in $S(L)$ (the bottom right figure~\ref{qL-SL}), since the fast oscillations of $q^*$ are in the ``anti-phase'' with those of the Lagrange multipliers $T^*$ (not shown in the figure), yielding a smooth increase of the entropy with a noticeable jump at $L_{\rm trans}$ only. Overall, \textbf{option 2} seems to be much more reliable \cite{Lenzi2000,preprint} for a description of the $q^*(L)$ behaviour almost everywhere except the domain ``small $M$ -- large $L$''. Nevertheless, we believe that a detailed study of peculiarities dealt with  \textbf{option 1} can also shed additional light upon the behaviour of the system described by the R\'{e}nyi/Tsallis parastatistics.

\begin{figure}[htb]
	\centerline{\includegraphics[height=0.32\textwidth,width=0.42\textwidth]{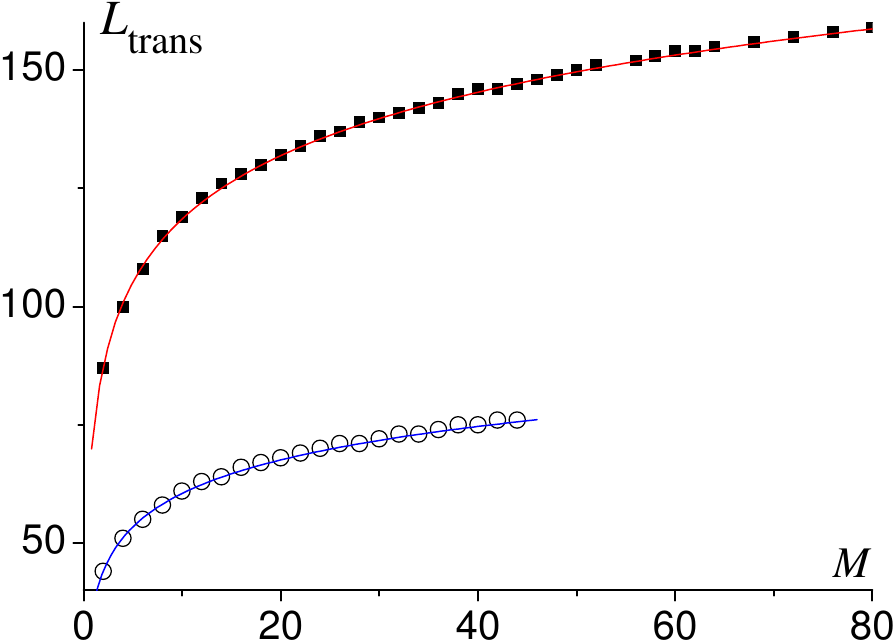}
				\hspace{1cm}
				\includegraphics[height=0.32\textwidth,width=0.42\textwidth]{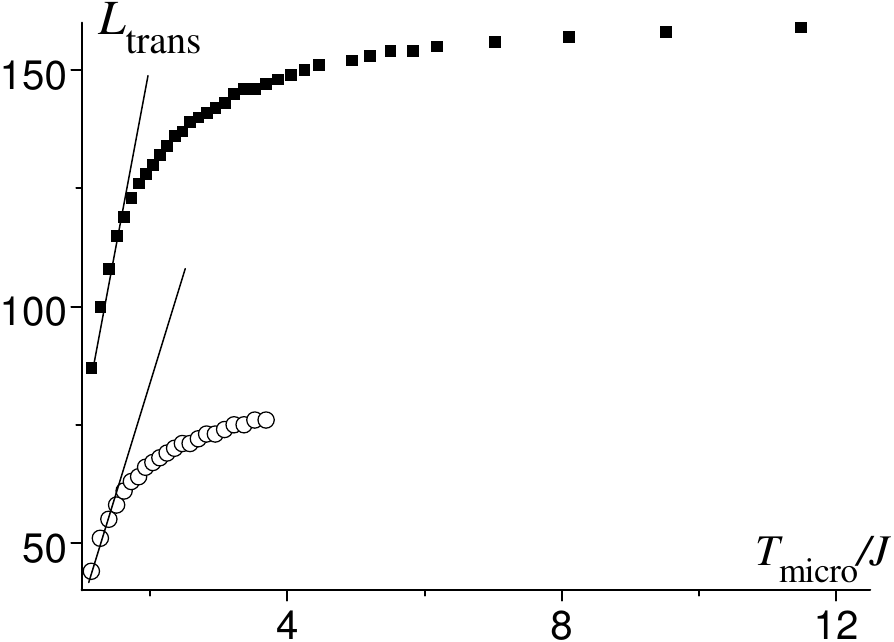}}
	\caption{(Colour online) Dependences of the transition length $L_{\rm trans}$ 
	on the number of pairs of the oppositely directed neighbouring spins~$M$ or the corresponding $T_{\rm micro}$, calculated using equation~(\ref{Tmicro}), for $N=200$ ($\blacksquare$) and $N=100$ ($\bigcirc$). 
		Red and blue lines: $L_{\rm trans}=A+B\ln M$ with $A=74$, $B=44.5$ and $A=37$, $B=23.5$, respectively.}
	\label{ol}	
\end{figure}

{We repeated the above calculations of the  R\'{e}nyi index $q^*$ and R\'{e}nyi entropy $S$  for $N=100$. Their dependences on the subsystem size $L$ were found to be analogous to those for $N=200$. However, the subsystem length $L_{\rm trans}$, at which the jump of the R\'{e}nyi index occurs, significantly decreases.

In figure~\ref{ol}, we plot the dependences of the  $L_{\rm trans}$ on $M$ for $N=100$ and  $N=200$. They are found to be approximated quite well by the logarithmic function $A+B\ln M$ with $A$ and $B$ being almost exactly twice larger for $N=200$ than for $N=100$.} In the right-hand panel, $L_{\rm trans}$ are replotted as functions of $T_{\rm ph}=T_{\rm micro}$ using the expression (\ref{Tmicro}):  the jump length grows linearly at low temperatures and barely changes at high temperatures. Unfortunately, no useful analytical results could be obtained for $L_{\rm trans}$; {the numerical calculations suggest that it should be proportional to $N$.}

\begin{figure}[h]
	\centerline{\includegraphics[height=0.32\textwidth,width=0.42\textwidth]{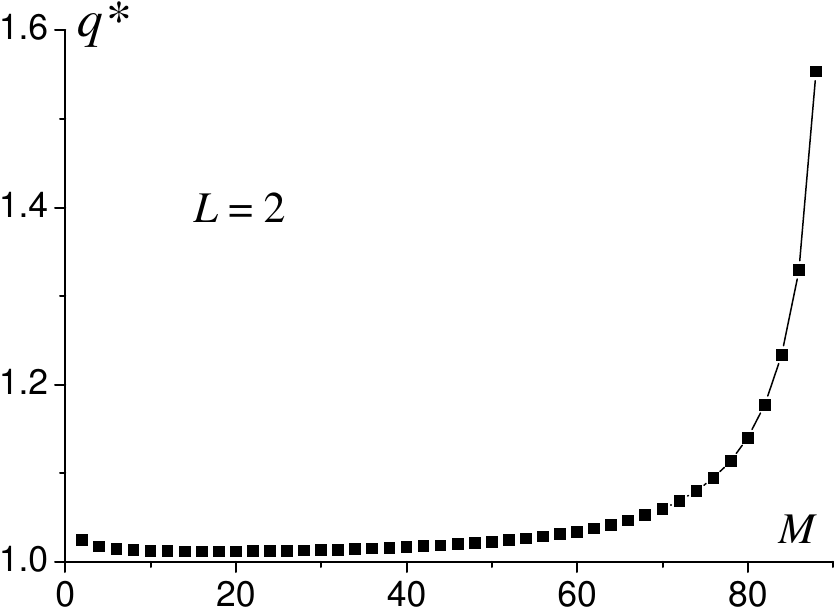}
		\hspace{1cm}
		\includegraphics[height=0.32\textwidth,width=0.42\textwidth]{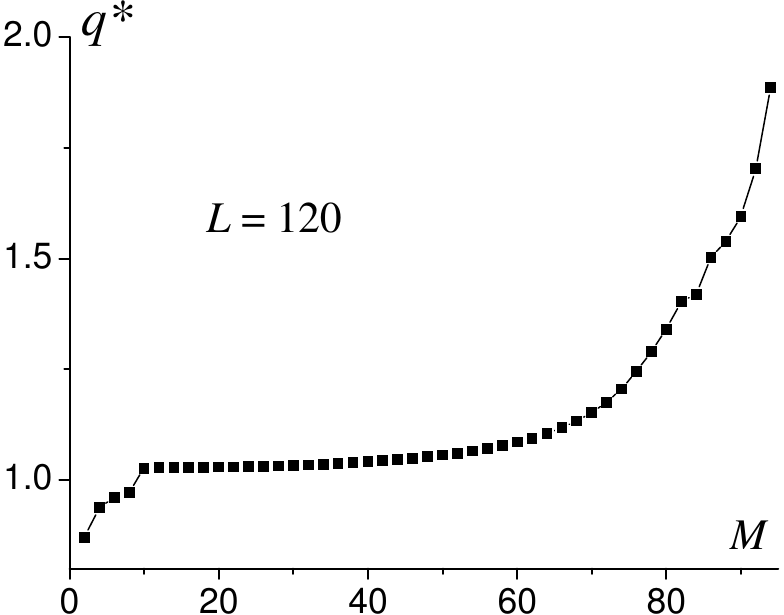}}
	\caption{Dependence of the R\'{e}nyi index $q^*$ on $M$ for $L=2$ (left-hand) and $L=120$ (right-hand). $N=200$. {The lines are guides to the eye.}}
	\label{qM}
\end{figure}

\begin{figure}[h]
	\centerline{\includegraphics[width=0.4\textwidth]{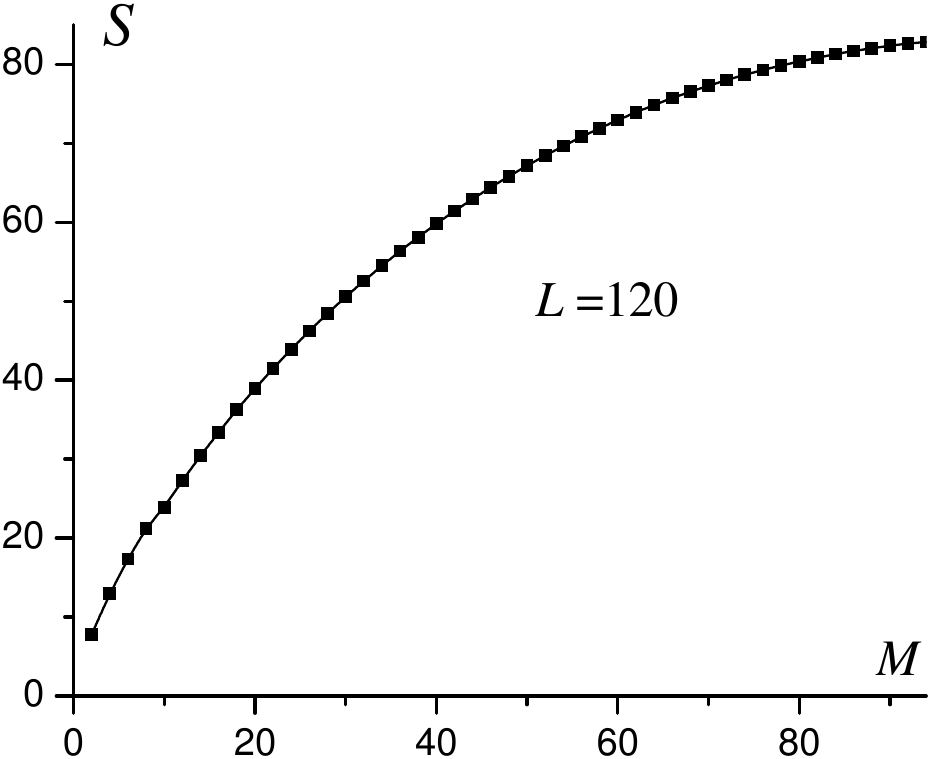}
		\hspace{1cm}
		\includegraphics[width=0.4\textwidth]{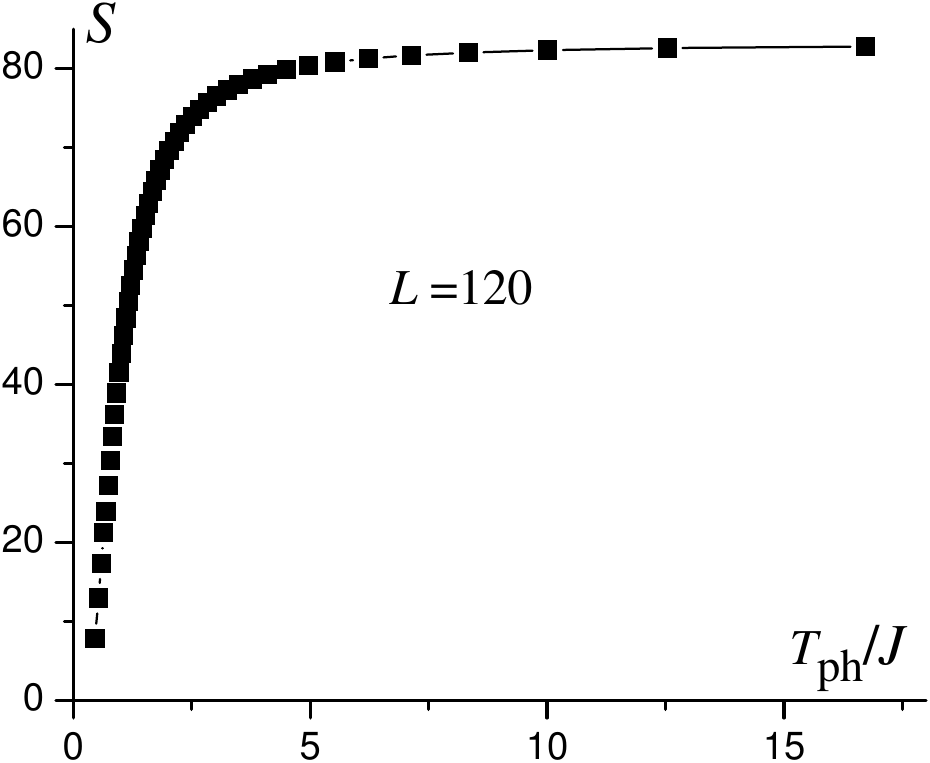}}
	\caption{Dependence of the R\'{e}nyi entropy on the number of
		pairs of the  oppositely directed neighbouring spins  $M$ (left) and temperature $T_{\rm ph}=T_{\rm micro}$ (right-hand) at constant subsystem length $L=120$. $N=200$.   {The lines are guides to the eye.}}
	\label{SL120}
\end{figure}

In figure~\ref{qM}, we plot the dependence of the R\'{e}nyi index $q^*$ on the number of pairs of the oppositely directed neighbouring spins $M$ at different $L$. For $L=2$, the $q^*(M)$ dependence is quite smooth. For $L=120$, a noticeable jump of the R\'{e}nyi index occurs at $M=12$, in agreement with the data shown in figure~\ref{ol}. Nevertheless, for the same $L=120$, no irregularity is observed in the dependence of the R\'{e}nyi entropy either on $M$ or on $T_{\rm ph}$ (figure~\ref{SL120}).
This result is quite interesting, since it shows that the conventional phase transitions in the 1D Ising model are absent not only in the limit $q\to 1$ but also at other values of the R\'{e}nyi indices.

\subsection{Problem II}
In the final part of this section, we consider a problem similar to that examined in \cite{preprint}, where the thermodynamics of the Ising spins chain was studied within the Tsallis statistics. As it has already been mentioned, the value of the $q$ index there was uniquely chosen by selection of the environment, which was composed of the ideal gas molecules interacting with the Ising spins.

Let us take one pair $(q^*, T^*)$ of solutions of equations~(\ref{selfC-main}), which was previously obtained for some generally arbitrary values $M$ and $L$. Then we put $q=q^*$ at all temperatures, whereas the Lagrange parameter $T$ is allowed to vary in quite a broad domain. 
The variation of $T$ means that the number of pairs of the  oppositely directed neighbouring spins $M$ in the entire chain varies too. We find $M$ and the internal energy $U$  by solving the self-consistency equation (\ref{selfUIsing}) as well as either the exact equation $T_{\rm ph}= T_{\rm micro}$, see 
(\ref{selfC-main}a) or its approximation $q T\approx T_{\rm micro}$. To do that we assume that $M$ could be a real number, and then we round the found solution to the nearest even integer. This rounded value is used  to determine the boundaries of the domains $[K_{\rm min},K_{\rm max}]$ of the allowed microstates at each temperature $T$. Again, of the multiple solutions at each given $T$ we choose that solution which corresponds to the largest  entropy.\footnote{The described algorithm means that we are looking for the values of $U$ that do not differ too much from those obtained using the microcanonical distribution. In fact, one can apply less strict conditions for the microstates numbers $K$, just demanding from them to ensure the positivity of the expression in the parentheses of equation~(\ref{pRIsing}). Obviously, a comparison of the obtained results with the observables calculated within the microcanonical ensemble loses its meaning in such a case.}

\begin{figure}[h]
	\centerline{\includegraphics[height=0.32\textwidth,width=0.42\textwidth]{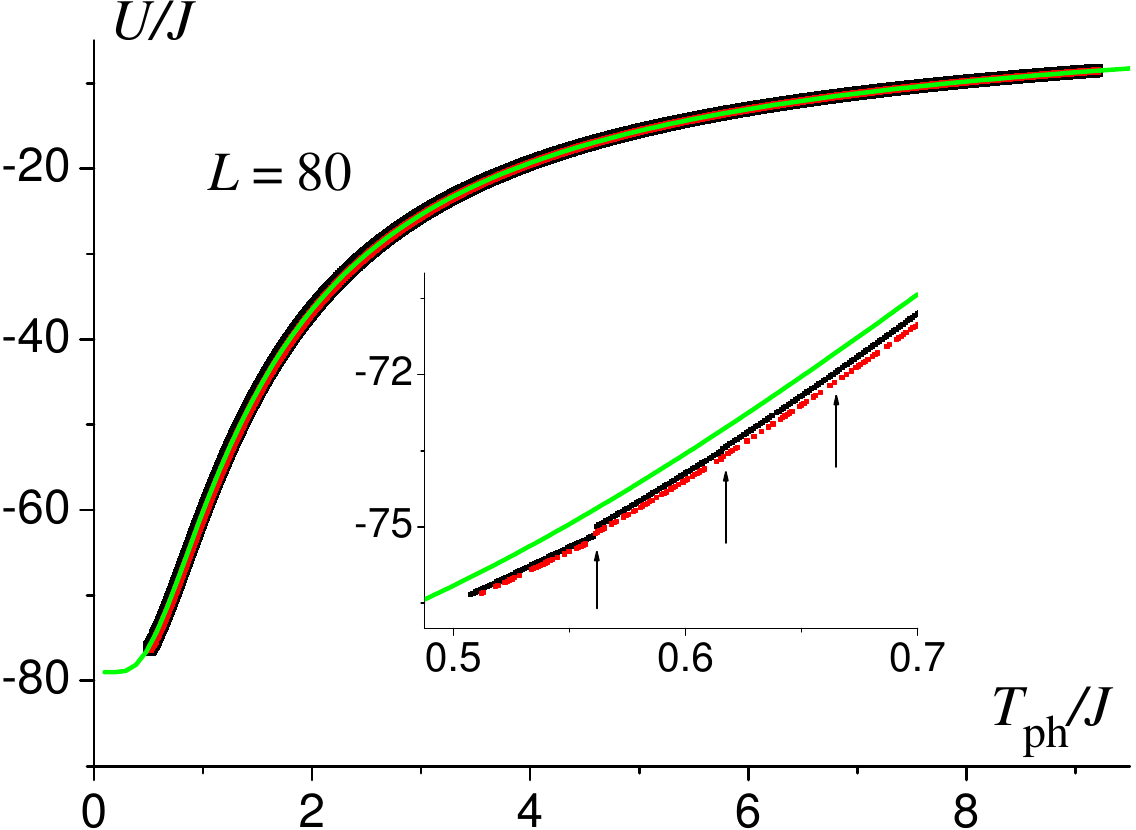}
		\hspace{1cm}
		\includegraphics[height=0.32\textwidth,width=0.42\textwidth]{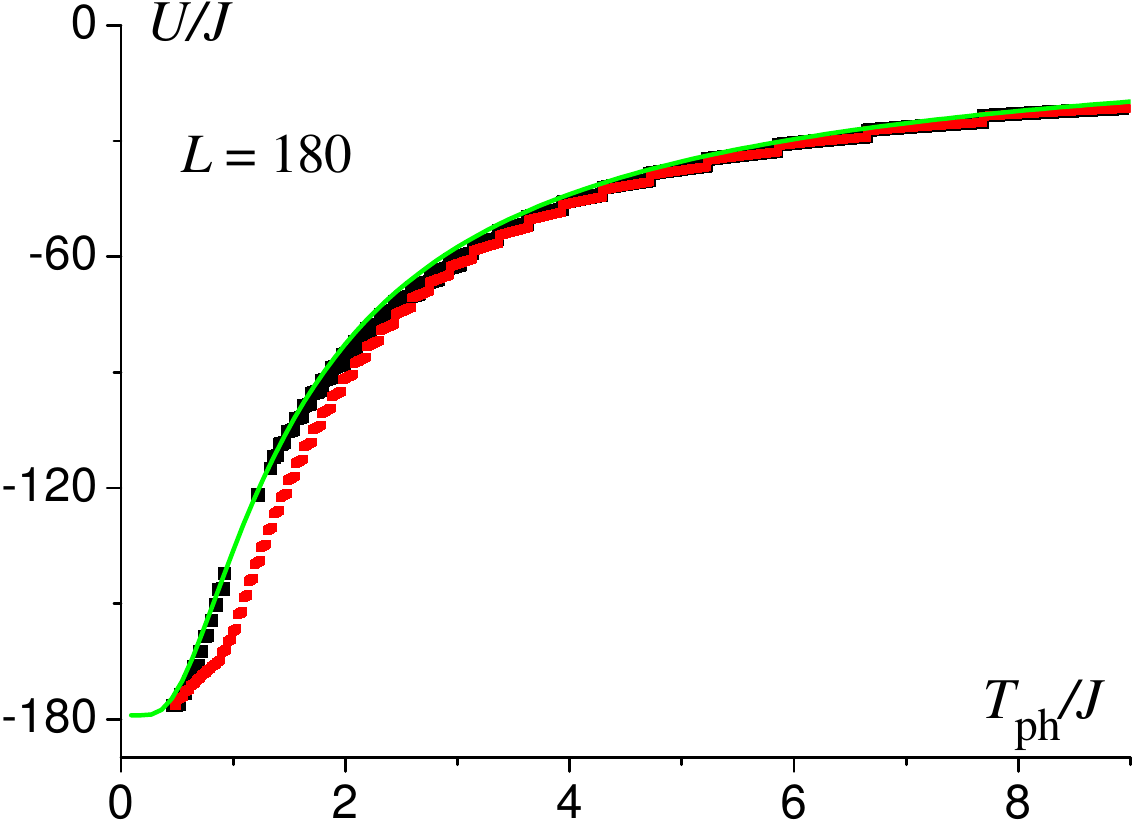}}
	\caption{(Colour online) The temperature dependence of the R\'{e}nyi internal energy. The calculations are carried out with constant $q=1.0219$ (left) and $q=0.94418$ (right);  $N=200$.
		The black and red points are found by using the relations $T_{\rm ph}= T_{\rm micro}$ and $q T\approx T_{\rm micro}$, respectively. The green lines are the internal energy of the canonical distribution $U^{(G)}=-J(L-1)\tanh(J/T_{\rm ph})$. The arrows in the inset indicate the jumps of $U$.  {(Only the lowest jump in the inset is discernible, while the magnitudes of two others are smaller then the size of the symbols)}. }
	\label{UTph}
\end{figure}

In figure~\ref{UTph}, we plot the  dependence of the internal energy of the subsystem on the physical temperature~(\ref{TR})  for two different $L=80$ (left-hand) and $L=180$ (right-hand) for $N=200$. We choose $M=30$ to fix the pair of solutions $(q^*, T^*)$ found for Problem~I, namely: $q^*(30,80)=1.0219$, $T^*(30,80)/J=1.1472$, and  $q^*(30,180)=0.94418$, $T^*(30,180)/J=2.3545$.
As one can see, the calculated exact dependences quite accurately follow the
$U^{(\text{G})}(T_{\rm ph})$ curves of the canonical distribution. It would be also true for the microcanonical ensemble: the internal energy defined by equation~(\ref{ESmean}) differs from that determined by the second equation in (\ref{CvG}) only by terms of the order $1/N$, which are quite unessential for a long enough Ising spin chain with $N=200$.

However, on a small scale, the dependences of the R\'{e}nyi internal energies on $T_{\rm ph}$  have a discrete stairway structure with steps and gaps  both in $U$ and in $T_{\rm ph}$. This structure is quite apparent for a larger subsystem (right-hand panel in figure~\ref{UTph}, $L=180$) and barely discernible and only at very low temperature for a significantly smaller subsystem (left-hand panel and the inset in figure~\ref{UTph}, $L=80$).
The steps become lower and wider with increasing $T_{\rm ph}$.
This stairway structure of the $U(T_{\rm ph})$ dependence is caused by the discreteness of the system, in particular of the allowed values of $M$ (even integers), $K$ (integers), and of the boundaries $[K_{\rm min}, K_{\rm max}]$ limiting the values of $K$, all of which change stepwise with the increasing temperature.
The increase of temperature and of $M$ is accompanied by the expansion of the interval $[K_{\rm min},K_{\rm max}]$ or even to its shift to the right. Physically, it means that more and more microstates are involved in the formation of the system thermodynamics.  This effect is discussed  more in detail in the appendix.

Note also that when the exact relation $T_{\rm ph}=T_{\rm micro}$ (\ref{selfC-main}a) or $q T\approx T_{\rm micro}$ is used in calculations, the $U(T_{\rm ph})$ of a smaller subsystem ($L=80$) is the same (left-hand panel). For a larger subsystem  ($L=180$, right-hand panel), such two curves coincide only at moderate to high temperatures, whereas at low temperatures they apparently deviate from each other. The exact curve follows the line of the internal energy of the canonical distribution. The big visible gap in the exact $U(T_{\rm ph})$ curve for $L=180$ at about $T_{\rm ph}\sim J$ is caused by the absence of solutions for quite a strong requirement $T_{\rm ph}=T_{\rm micro}$ and probably does not bear any physical significance; this gap can vanish if one loosens the above requirement.   
\begin{figure}[h]
	\centerline{\includegraphics[height=0.32\textwidth,width=0.42\textwidth]{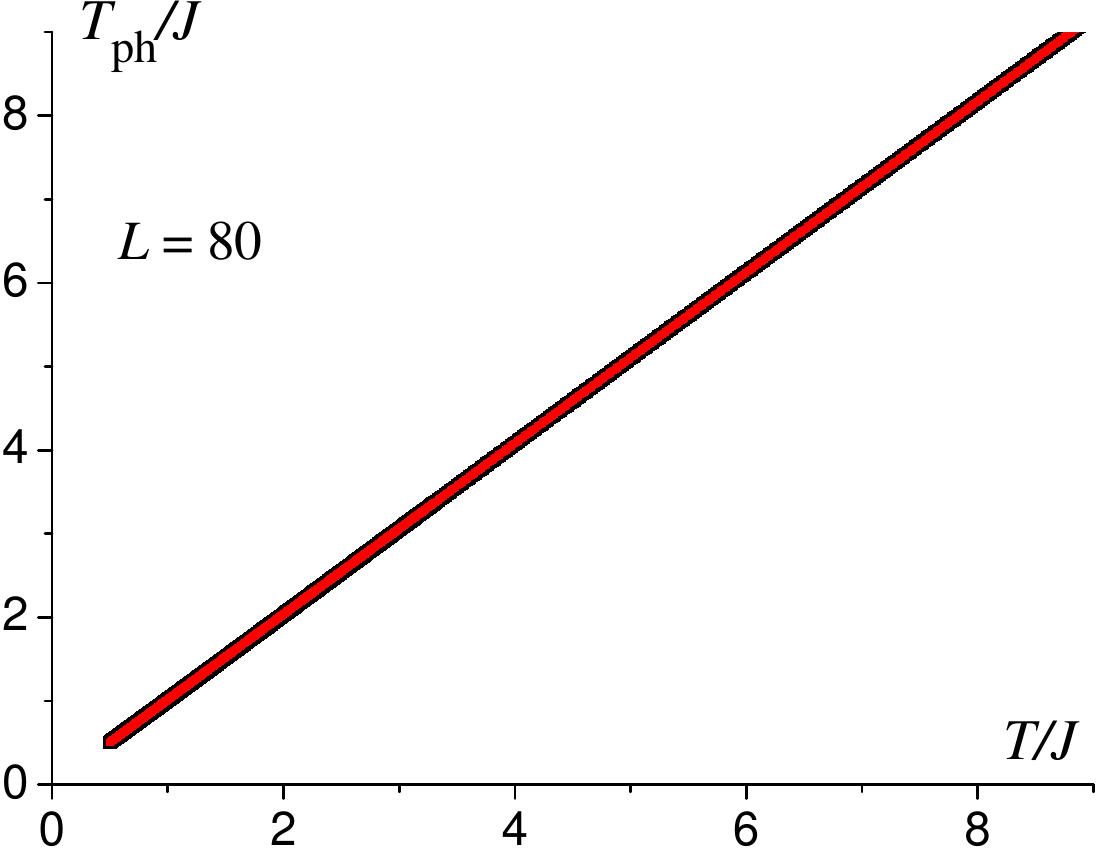}
		\hspace{1cm}
		\includegraphics[height=0.32\textwidth,width=0.42\textwidth]{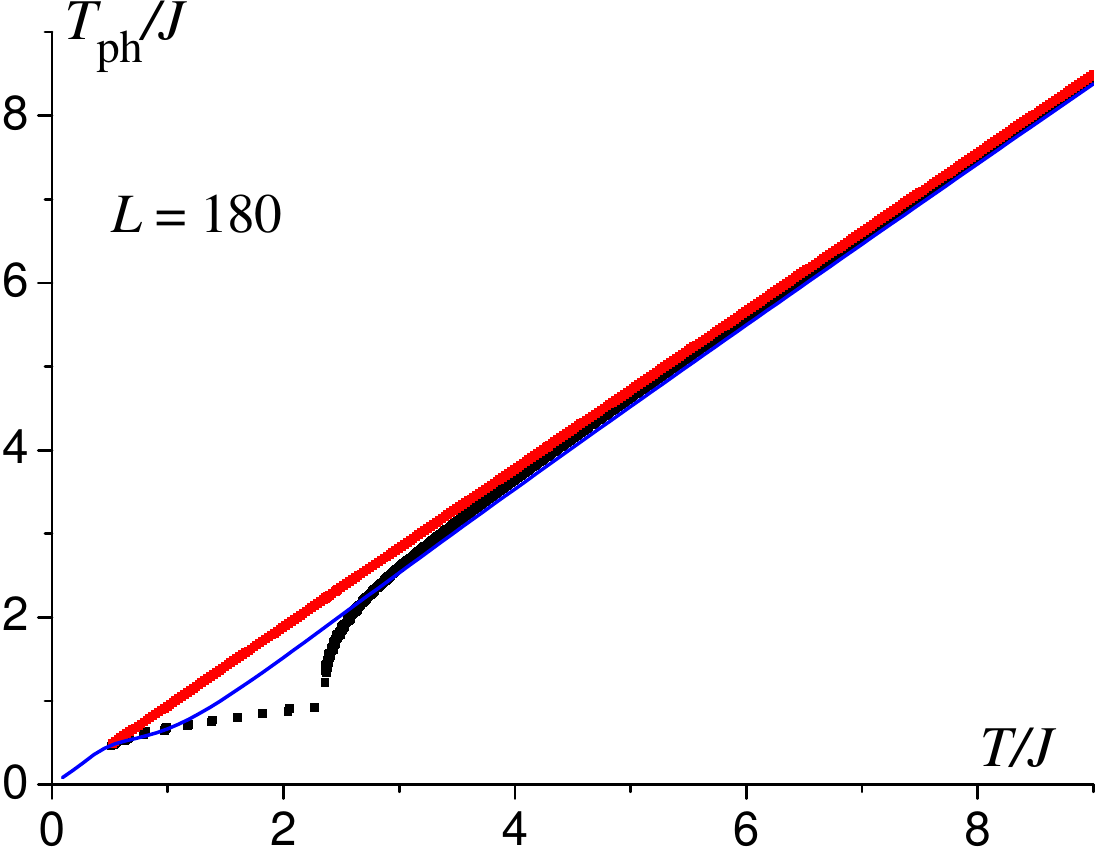}}
		\caption{(Colour online) The dependence $T_{\rm ph}(T)$. The calculations are carried out with constant $q=1.0219$ (left-hand panel) and $q=0.94418$ (right-hand panel).  
		The black and red symbols are calculated by using the exact $T_{\rm ph}= T_{\rm micro}$ and approximate $q T\approx T_{\rm micro}$  relations, respectively. {The blue line is found from equation~(\ref{TRapprox}). $N=200$.}}
	\label{Tph(T)} 
\end{figure}

{Some light on the origin of the peculiarities of the presented above $U(T_{\rm ph})$ dependences is shed by the corresponding $T_{\rm ph}(T)$ dependences, plotted in figure~\ref{Tph(T)}}. The two $U(T_{\rm ph})$ curves in the left-hand panel of figure~\ref{UTph} overlap because the corresponding $T_{\rm ph}(T)$ curves coincide: regardless of the relation used for calculation of the allowed microstates, $T_{\rm ph}$ for $L=80$ follows the same straight line $q T$. This agrees with equation~(\ref{TRapprox}), according to which $T_{\rm ph}=qT$ in the first approximation, whereas the next term in the expansion series is proportional to $(1-q)^2 (L-1)$ and is negligible, when the R\'{e}nyi index only slightly deviates from unity.

\begin{figure}[h]
	\centerline{\includegraphics[height=0.28\textwidth,width=0.38\textwidth]{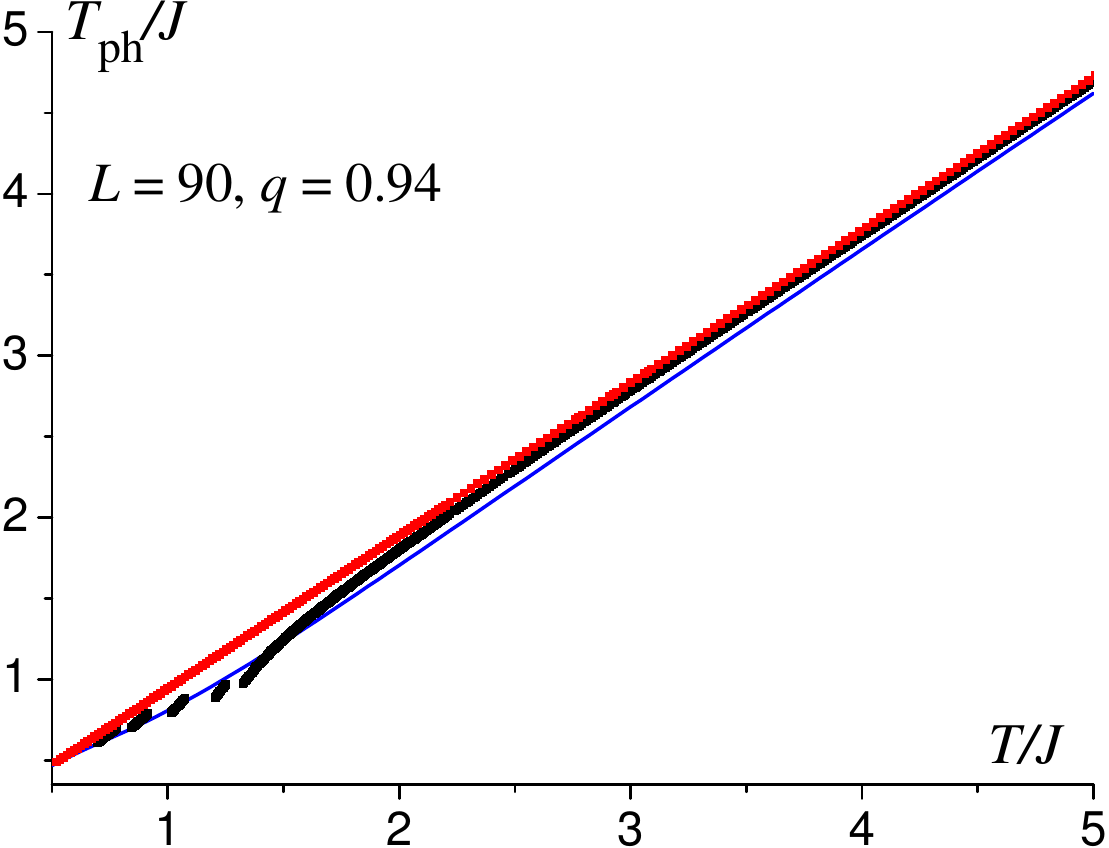}
		\hspace{1cm}
		\includegraphics[height=0.28\textwidth,width=0.38\textwidth]{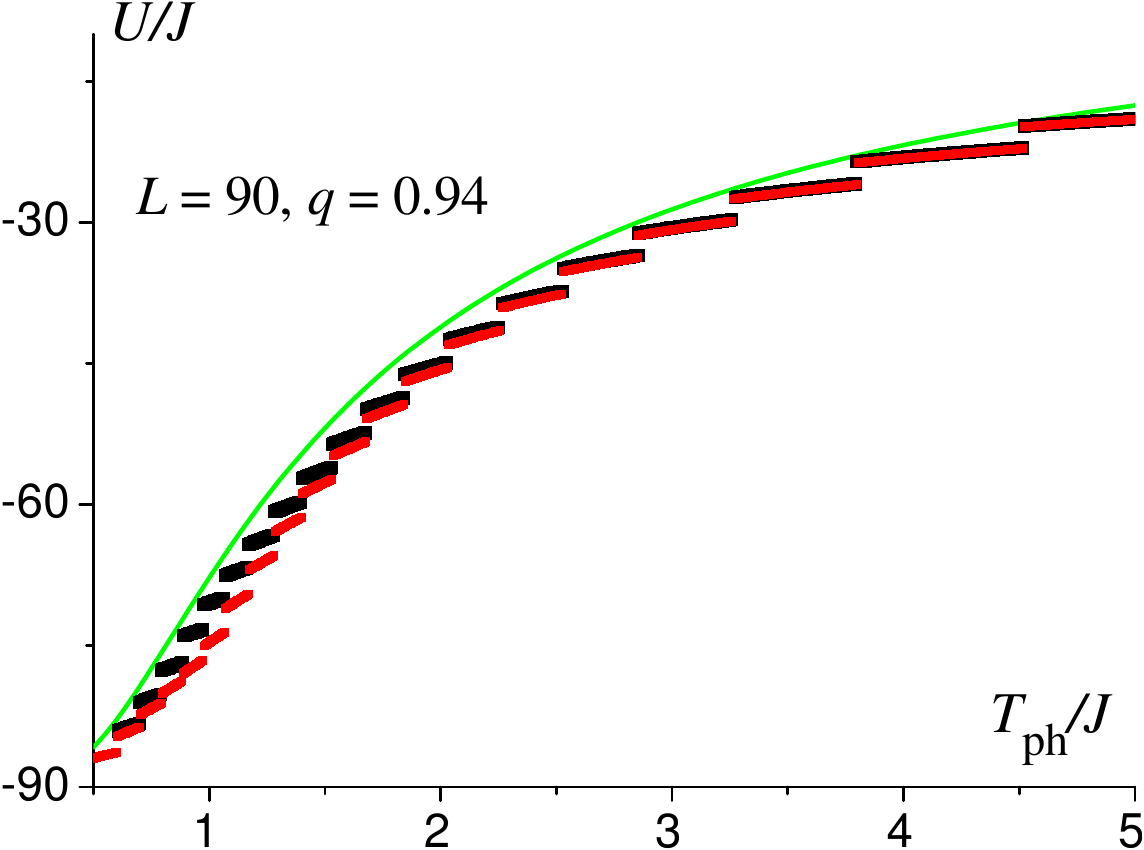}}
	\caption{(Colour online) The dependences $T_{\rm ph}(T)$ and $U(T_{\rm ph})$ for $N=100$ and $L=90$; $q=0.94418$.  
		The symbols and lines are the same as in Figs.~\ref{UTph} and \ref{Tph(T)}.}
	\label{Tph(T)N100}
\end{figure}

{
The situation changes for a  larger subsystem, $L=180$ (the right-hand panel in figure~\ref{Tph(T)}).
In this case, the difference between the chosen value of the R\'{e}nyi index, $q=0.94418$, and unity is greater and the product $(1-q)^2 (L-1)$ is much greater. Nevertheless, the  dependence $T_{\rm ph}(T)$ is still linear at very low and at moderate to high $T$, which means that the second term in equation~(\ref{TRapprox}) becomes negligible at these temperatures, anyway. The approachment of the limits $T\to 0$ and $T\to \infty$ by the calculated $T_{\rm ph}(T)$ dependence is well approximated by equation~(\ref{TRapprox}), as indicated by the blue line in figure~\ref{Tph(T)}. At intermediate $T$, however, $T_{\rm ph}(T)$ calculated with the exact relation $T_{\rm ph}= T_{\rm micro}$ essentially deviates both from the straight line $q T$ and from the curve, corresponding to equation~(\ref{TRapprox}). This temperature range clearly has two domains with different character of
the $T_{\rm ph}(T)$ dependence. It is the deviation of $T_{\rm ph}(T)$ from $T_{\rm ph}=qT$ that causes the difference between the corresponding $U(T_{\rm ph})$ dependences (the black and red symbols in figure~\ref{UTph}). Obviously, the difference between the exact $T_{\rm ph}(T)$ and calculated from equation~(\ref{TRapprox}), means that higher terms in the expansion of equation (\ref{TR}) over $q-1$ must be taken into account, because the quadratic term only included in equation~(\ref{TRapprox}) is not sufficient at the intermediate temperatures.

In figure~\ref{Tph(T)N100} we plot the $T_{\rm ph}(T)$ and $U(T_{\rm ph})$ dependences for a twice smaller entire system, $N=100$, for
$L=90$ and $q=0.94418$, i.e., the same $L/N$ ratio (a twice smaller subsystem) and $q$ as in the right-hand panels of figures~\ref{UTph}, \ref{Tph(T)}. Notable is a larger deviation of the calculated $U(T_{\rm ph})$ dependence from the canonical curve, which is totally expected. The exact $T_{\rm ph}(T)$ dependence, on the other hand, is much better approximated by equation~(\ref{TRapprox}), as well as it is much closer to the line $T_{\rm ph}=qT$, which is also expected for smaller $L$ at the same value of $q$. The stairway structure of the shown dependences is more pronounced, and the step width is increased, as compared to the case $N=200$, in agreement with equation~(\ref{stepwidth}); see the discussion in appendix.}

Worth mentioning is the monotonous dependence of the physical temperature on $T$ { for the above considered values of $q$}. In \cite{in-preprint17}, the non-monotonous behaviour $T_{\rm ph}(T)$ of the $8$-level system with an equidistant discrete spectrum in the Tsallis statistics was observed and attributed to the metastable states. In our case, no metastable states have been detected, as it is explained in detail in the Appendix. {Interestingly, as preliminary calculations show, the $T_{\rm ph}(T)$ dependence for $q$, much different from~1, becomes non-monotonous at low temperatures, but the $U(T_{\rm ph})$ dependence remains monotonous, and no metastable states are still detected. We shall address these findings more in detail elsewhere.}

\section{Conclusions and outlook}\label{sec6}

In this paper, we study the behaviour of the 1D Ising model in the framework of the R\'{e}nyi statistics. We consider the R\'{e}nyi statistics as a certain alternative to the microcanonical distribution in describing the size effects for the subsystem, which is supposed to be comparable with its surroundings or even greater than the environment.

First, we consider Problem I, based on our assumption that the observables (the internal energy and the physical temperature) are the same in both statistics. This allowed us to self-consistently calculate the R\'{e}nyi index $q$ and to study its evolution from the values slightly greater than unity, when the subsystem size $L$ is much less the that of the total system $N$, through the jump of $q$ at the intermediate values of $L$, and up to the limit $q\to 0$, when the size of environment is vanishing, and the system is governed by the microcanonical distribution. A downward jump of the R\'{e}nyi index at $L=L_{\rm trans}$ is accompanied by a notable change of the system entropy, which gives us reasons to associate it with the entropic phase transition. At the same time, the curve describing the temperature behaviour of the R\'{e}nyi entropy has no anomalies, which indicates the absence of conventional phase transitions.
The derivative of the R\'{e}nyi entropy with respect to the order parameter at $\eta\to 0$ turns out to be proportional to the heat capacity $C_V$ of the subsystem, as it is expected from the basic thermodynamic relations. The calculated behavior of the R\'{e}nyi index $q$ is found to qualitatively agree with the one given by the formulae, which follow from the simple thermo-fluctuation approach \cite{Bashkirov-Book} or more sophisticated statistical methods \cite{Almeida} and invoke an inverse heat capacity of the environment.

Problem~II considered is dealt with the temperature behaviour of the internal energy. To this end, we have loosened the first assumption $U_R=U_{\rm micro}$ made earlier. {We use fixed values of the R\'{e}nyi index $q$ (which do not substantially differ from the unity)}, while the allowed microstates are determined from equation~(\ref{selfC-main}a).
We obtain a monotonous temperature behaviour of $U(T_{\rm ph})$ with no metastable states detected. On a small scale,  $U(T_{\rm ph})$, $T_{\rm ph}(T)$, and $M(T_{\rm ph})$ exhibit stairway structures, which we attribute to the discreteness of the system spectra, i.e., those are the size effects that disappear in the thermodynamic limit. These size effects become stronger when the total particle number $N$ decreases, and the {difference between $q$ and unity increases}, as the subsystem particle number $L$ increases at $N=$~const. 

The approach similar to ours had been adopted in \cite{fin-bath}, when a portion of the isotropic XY spin chain was interpreted as the subsystem, whereas the remaining spins composed the bath. Though the statistics of the overall system was taken to be Gibbsian (a concept of the superbath was introduced, and the entire spin chain was immersed in the infinite environment), some thermodynamic anomalies, obviously caused by the finite bath effects, were observed. It could be interesting to follow this line of investigation, namely, to loosen the requirement for $K$, demanding only ``unphysical'' states avoidance, and to study the temperature behaviour of the same thermodynamic functions as in \cite{fin-bath}.

Another problem worth studying is a generalization of our approach to the two-temperature formalism by introducing the concept of the system and bath temperatures, see \cite{Ramshaw} and the references therein. In the framework of the thermo-fluctuation approach and the related methods \cite{Bashkirov-Book,PhysA2019}, the ``system'' temperature $T$ is considered to be a random quantity, while the ``bath'' temperature $T_0=q T$ coincides with the mean value of the stochastic variable $T$. It does not allow one to introduce  the mentioned two-temperature concept reliably.
At the same time, within the purely statistical approaches \cite{Almeida} supplemented by the thermodynamic consideration \cite{Klimontovich,24in-Ramshaw}, the two-temperature formalism becomes quite realistic, at least, if the heat bath by size and energy content is greater than or comparable to the system under consideration~\cite{Ramshaw}. 
However, a concept of the ``higher'' bath temperature $T_{\rm bath}=q T$ at $q>1$ cannot be directly adopted in our case, since the R\'{e}nyi indices remain greater than unity for large enough subsystems (up to $L_{\rm trans}$, see figure \ref{ol}), when the environment length $N-L_{\rm trans}$ becomes smaller than that of the system. The assumption made on page~\pageref{assumption} should be re-examined if one is going to directly introduce the two-temperature formalism using, for instance, the exact statistical relations \cite{Almeida} or the second law of  thermodynamics \cite{Ramshaw}.

All the above mentioned problems can be the subject of future investigations.

\section*{Acknowledgement}
The authors are grateful to Prof. A. Rovenchak and Dr. V. Pastukhov from the chair of theoretical physics of the Ivan Franko National University of Lviv for the interest in this work, encouragement and stimulating discussions.

\section*{Appendix}
\renewcommand{\theequation}{A.\arabic{equation}}
\setcounter{equation}{0}

As it has been discussed in section~\ref{sec5}, the overall stairway structure of the $U(T_{\rm ph})$ dependence is caused by the 
jumps in the number of the pairs of the oppositely directed neighbouring spins $M$ with increasing temperature. The qualitative physical picture shows that the jumps occur when this increase is sufficient to flip yet another spin and create (for ferromagnetic $J$) two extra such pairs, so that the next energy level could be populated. Technically, this happens whenever the actual solution $M$ (a real number) of the system increases through an odd number, say $2n+1$. Just below the jump it is rounded to the nearest even integer number $2n$, and just above the jump to $2n+2$.

\begin{figure}[h]
	\centerline{\includegraphics[height=0.4\textwidth]{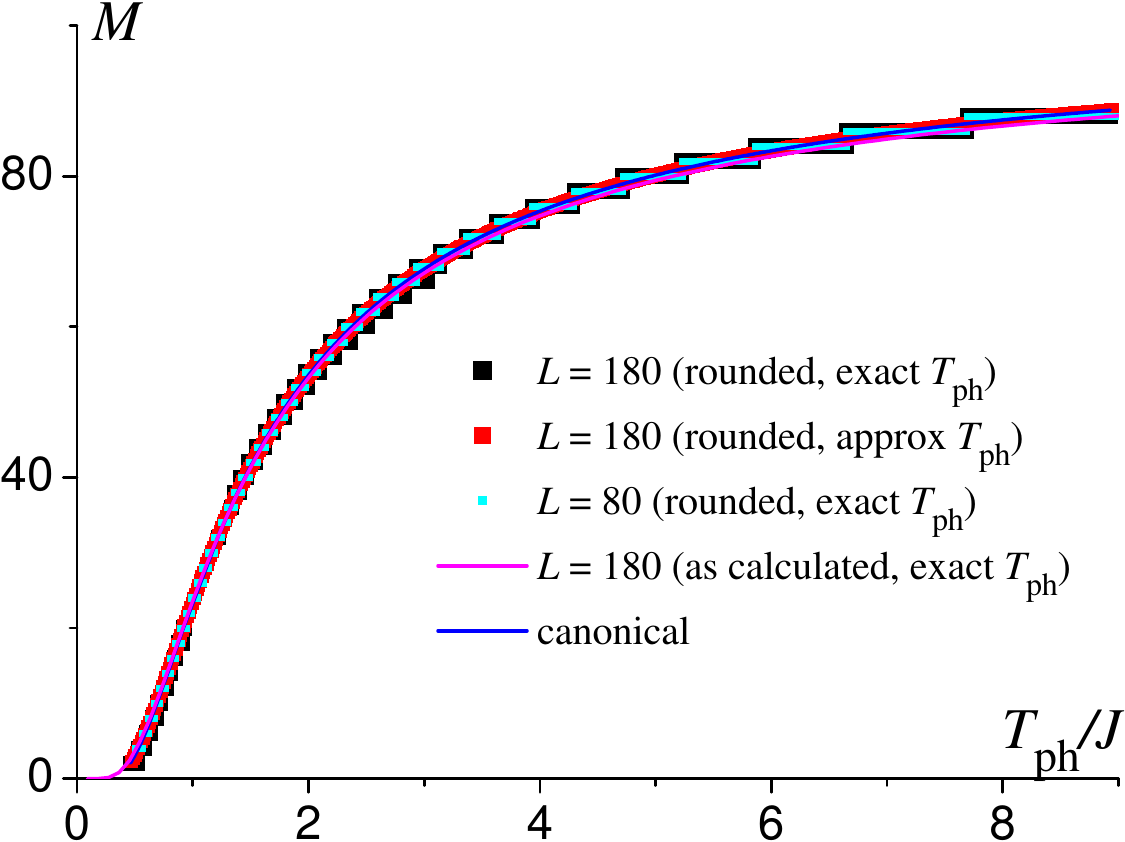}}
	\caption{(Colour online) The temperature dependence of the number $M$ of the pairs of the oppositely directed neighbouring spins
		on temperature $T_{\rm ph}$, calculated for $L=80$ and $L=180$ with the exact $T_{\rm ph}=T_{\rm micro}$ and approximate $qT=T_{\rm micro}$ equations, the real number solution as calculated, or rounded to the nearest even integer, and, finally, obtained from the canonical distribution, equation~(\ref{MT}). $N=200$.}
	\label{MTph}
\end{figure}

As seen in figure~\ref{MTph}, in contrast to the $T_{\rm ph}(T)$ and $U(T_{\rm ph})$ curves, the dependence $M(T_{\rm ph})$ is basically the same for different $L$ and does not depend on the employed approximation for $T_{\rm ph}$. If one uses the canonical distribution equation for $M$, equation~(\ref{Theta}), as a first approximation, then
\begin{equation}
	\label{MT}
	M=\frac{N}2\left(1-\tanh\left[\frac{J}{T_{\rm ph}}\right]\right).
\end{equation}
Here, $M$ is independent of $L$ and is determined by the size of the entire system $N$, as expected. The calculated with the R\'{e}nyi distribution $M$ follows the canonical $M(T_{\rm  ph})$ curve of equation~(\ref{MT}), as seen in figure~\ref{MTph}. The  width of a step in the stairway structure of $M$ (and thereby of $U$) can be estimated as the change of temperature $\Delta T_{\rm ph}$, required to increase $M$ by 2:
\begin{equation}\label{stepwidth}\Delta T_{\rm ph}\approx\frac{4T^2_{\rm ph}}{NJ}\cosh^2\left[\frac{J}{T_{\rm ph}}\right].\end{equation}
According to this, $\Delta T_{\rm ph}$ increases with temperature everywhere, except for the region of very low temperatures, $T_{\rm ph}/J<0.8$. The predicted behaviour agrees with the picture in figures~\ref{UTph}, \ref{MTph} both qualitatively and quantitativly. The step width also decreases with  increasing $N$ and tends to zero in the thermodynamic limit, yielding smooth temperature dependences of $M$ and $U$.

It should be mentioned that the nature of the observed in figures~\ref{UTph}, \ref{MTph} stairway structures is completely different from that of the steps depicted in figure~2 of \cite{preprint}, where they appeared due to the application of ``the Maxwell's law of areas'' when dealing with the metastable states. Besides, the steps of $U(T_{\rm ph})$ in the cited paper become less wide as  temperature increases. 

\begin{figure}[h]
	\centerline{\includegraphics[height=0.33\textwidth]{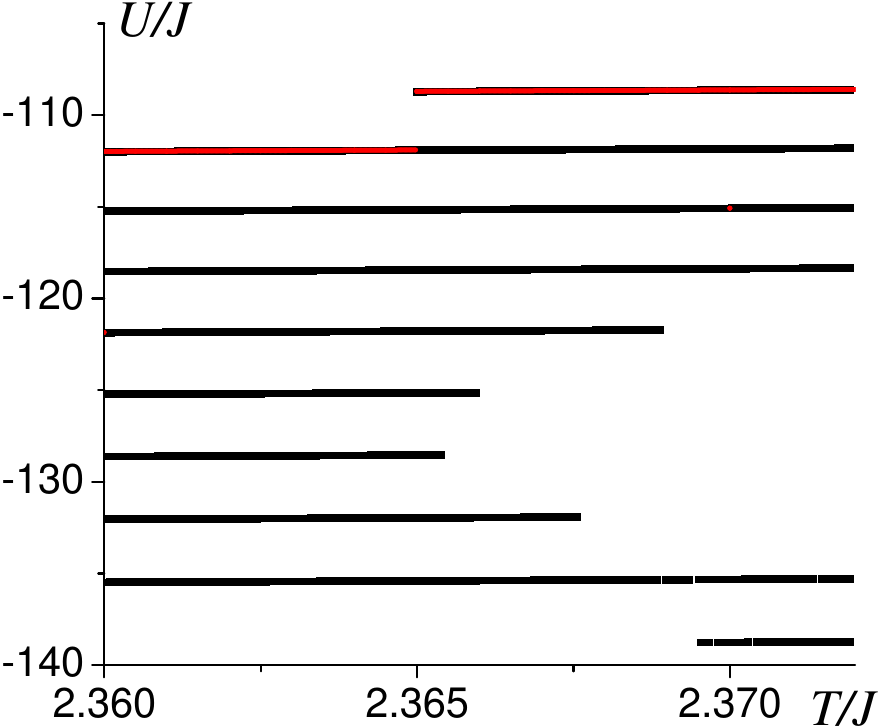}
		\hspace{1cm}
		\includegraphics[height=0.33\textwidth]{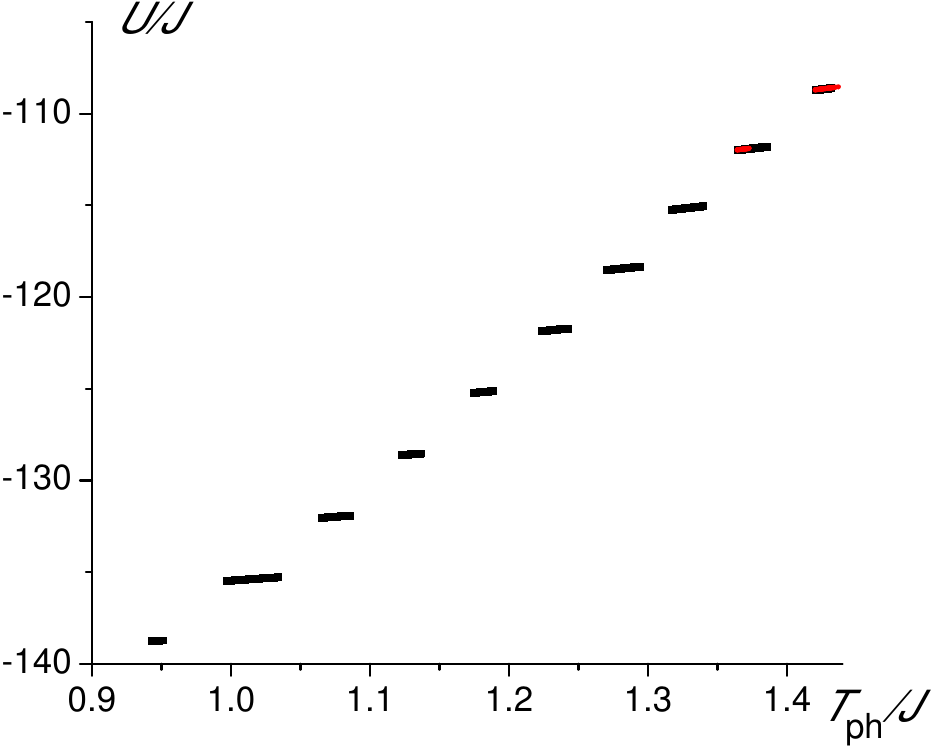}}
	\caption{(Colour online) The temperature dependences of solutions (\ref{selfUIsing}), (\ref{selfC-main}a) for the R\'{e}nyi internal energy. The calculations are carried out with $q=0.94418$, using the exact relation $T_{\rm ph}= T_{\rm micro}$ for $L=180$, $N=200$.  
		The black points represent all the found solutions, while the red ones are the adopted solutions, which correspond to the largest entropy. }
	\label{nometa}
\end{figure}

At least for the considered values of $L$, $N$, and $q$, we do not observe the metastable states like in \cite{preprint}. Even when the system of equations (\ref{selfUIsing}), (\ref{selfC-main}a) has multiple solutions with different $M$ at the same value of $T$, these solutions, when plotted as functions of $T_{\rm ph}$, do not overlap. Moreover, there exist gaps in the allowed values of $T_{\rm ph}$. The gaps are to be expected, since $T_{\rm ph}$ given by equation~(\ref{TR}) is a function of $M$ via $K_{\rm min}$ and $K_{\rm max}$, and these increase stepwise. This behaviour is illustrated in figure~\ref{nometa} for the vicinity of one of the jumps in $U$. Note that among these multiple solutions, the one with the largest $M$ always corresponds to the largest entropy and is adopted. A jump between the adopted
solutions (shown in red), as well as the gap in the values of $T_{\rm ph}$, are clearly seen in the figures.

\bibliographystyle{cmpj}


\ukrainianpart

\title{Модель Ізінга у статистиці Рені: ефекти скінченного розміру}
\author{В. В. Ігнатюк\refaddr{label1, label2}, А. П. Моїна\refaddr{label1}}
\addresses{
\addr{label1} Інститут фізики конденсованих систем НАН України, вул. Свенціцького, 1, 79011 Львів, Україна
\addr{label2} Львівський національний університет ім. Івана Франка, вул. Університетська, 1,  79007 Львів, Україна
}

\makeukrtitle

\begin{abstract}
\tolerance=3000%
Основні засади статистики Рені застосовано для опису ефектів скінченого розміру у одномірній моделі Ізінга. Внутрішня енергія та температура системи розрахована з використанням розподілу Рені та вважаються рівними аналогічним величинам, обчисленим у мікроканонічому ансамблі. Це дозволяє самоузгоджено розрахувати індекс Рені $q$ та параметр Лагранжа $T$, пов'язати їх з фізично спостережуваною температурою $T_{\rm ph}$ і показати, що у широкому діапазоні температур у системі можливі ентропійні фазові переходи. Також досліджено температурну залежність внутрішньої енергії $U(T_{\rm ph})$ при фіксованому $q$ та вплив ефектів скінченого розміру на термодинаміку системи.

\keywords {статистика Рені, мікроканонічний ансамбль, ентропійні фазові переходи, модель Ізінга}

\end{abstract}

\end{document}